\documentclass{nature}
\usepackage{color}
\usepackage{amssymb}
\usepackage{acronym}
\usepackage{amsmath}
\usepackage{hyperref}
\usepackage{graphicx}
\usepackage{lineno}
\usepackage{bm}
\usepackage{multibib}
\newcites{supp}{Supplementary References}

\def\IncludeSM{1}

\hypersetup{
    pdfnewwindow=true,      
    colorlinks=true,        
    linkcolor=blue,         
    citecolor=blue,         
    filecolor=blue,         
    urlcolor=blue           
}

\allowdisplaybreaks

\def\p{\partial}
\def\Lie{{\cal L}}
\def\Lam{\Lambda}
\def\blam{\bar{\lambda}}
\def\Lamt{\tilde{\Lambda}}
\def\dLamt{\tilde{\delta{\Lambda}}}
\def\non{\nonumber}                     
\def\half{\frac{1}{2}}
\def\e{{\rm e}}
\def\i{{\rm i}}
\def\gccm{g\,cm^{-3}}
\def\Msun{M_{\odot}}
\def\GMc2{G M_{\odot} c^{-2}}
\def\Mpc{Mpc}
\def\eps{\epsilon}
\def\sf{{\rm SF}}
\def\I{\mathcal{I}}
\def\M{\mathcal{M}}
\def\O{\mathcal{O}}
\def\vareps{\varepsilon}
\def\vrho{\varrho}
\def\rad{rad}
\def\lm{{\ell m}}
\def\teobLR{{t^{\rm EOB}_{\Omega \, \rm peak}}}
\def\teobpo{{t^{\rm EOB}_{\Omega_{\rm orb}^{\rm max}}}}
\def\tdomgmax{{t^{\rm NR}_{\dot{\omega}_{22}\, \rm peak}}}
\def\tAmax{{t_{A_{22}^{\rm max}}^{\rm NR}}}
\def\tnrextr{{t_{\rm extr}^{\rm NR}}}
\def\teobNQC{{t^{\rm EOB}_{\rm NQC}}}
\def\lm{{\ell m}}
\def\v{v_\varphi}
\def\non{\nonumber}                  
\def\half{\frac{1}{2}}
\def\de{\partial}
\def\lm{{\ell m}}
\def\g{{\gamma}}
\def\o{{\rm o}}
\def\ii{{\rm i}}
\def\l{{\ell }}
\def\r{{\hat{r}}}
\def\ph{\varphi}
\def\th{\vartheta}
\def\A{{\cal A}}
\def\I{{\cal I}}
\def\F{{\cal F}}
\def\J{{\cal J}}
\def\U{{\cal U}}
\def\E{{\cal E}}
\def\B{{\cal B}}
\def\M{{\cal M}}
\def\S{{\cal S}}
\def\O{{\cal O}}
\def\X{{\bf X}}
\def\Y{{\cal Y}}
\def\N{{\cal N}}
\def\W{{\cal W}}
\def\Z{{\cal Z}}
\def\R{{\cal R}}
\def\k{{\hat{\hat{k}}}}
\def\a{\alpha}
\def\b{\beta}
\newcommand\be{\begin{equation}}
\newcommand\ee{\end{equation}}
\def\hr{\hat{r}}
\def\ha{{\hat{a}}}
\def\rmd{{\rm d}}
\def\rcm{R_*}
\def\rhoo{S_\pm}
\def\rhoi{R_\pm}
\def\Msun{M_\odot}

\def\pphi{{p_{\varphi}^0}}
\def\CE{{\rm CE_0}}
\def\UE{{\rm UE_0}}

\def\TEOBResumS{\texttt{TEOBResumS}}
\def\TEOBResumSHyp{\texttt{TEOBResumSHyp}}
\def\TEOBResumSP{\texttt{TEOBResumSP}}
\def\TEOBResumShm{\texttt{TEOBiResumS\_SM}}
\def\TEOBResumSecce{\texttt{TEOBResumSecce}}
\def\TEOBResumROM{\texttt{TEOBResum\_ROM}}
\def\SEOBNRvq{{\texttt{SEOBNRv4}}}
\def\SEOBNRvqT{{\texttt{SEOBNRv4T}}}
\def\CC{{C\nolinebreak[4]\hspace{-.05em}\raisebox{.4ex}{\tiny\bf ++}}}

\newcommand{\an}[1]{{\textcolor{red}{\texttt{AN: #1}} }}
\newcommand{\pr}[1]{{\textcolor{cyan}{\texttt{PR: #1}} }}
\newcommand{\bs}[1]{{\textcolor{blue}{\texttt{SB: #1}} }}
\newcommand{\mb}[1]{{\textcolor{purple}{\texttt{MB: #1}} }}
\newcommand{\rg}[1]{{\textcolor{green}{\texttt{RG: #1}} }}
\newcommand{\gc}[1]{{\textcolor{violet}{\texttt{GC: #1}} }}
\usepackage[normalem]{ulem} 
\newcommand{\oldtxt}[1]{{\sout{#1}}}
\newcommand{\oldnewtxt}[2]{{\sout{#1}{\red{{#2}}}}}
\newcommand{\newtxt}[1]{{\textcolor{magenta}{{\bf[new]} #1}}}
\newcommand{\red}[1]{{\textcolor{red}{#1}}}

 \makeatletter
\let\saved@includegraphics\includegraphics
\AtBeginDocument{\let\includegraphics\saved@includegraphics}
\renewenvironment*{figure}{\@float{figure}}{\end@float}
\makeatother

 \title{GW190521 as a dynamical capture of two nonspinning black holes}
	
\author{R.~Gamba$^{1}$, M.~Breschi$^{1}$, G.~Carullo$^{1,2,3,7}$, S.~Albanesi$^{4,5}$, P.~Rettegno$^{4,5}$,  S.~Bernuzzi$^{1}$, A.~Nagar$^{5,6}$\footnote{Corresponding author: alessandro.nagar@gmail.com}}

\begin{document}
\maketitle

\begin{affiliations}
\item Theoretisch-Physikalisches Institut, Friedrich-Schiller-Universit{\"a}t Jena, Jena, 07743,Germany
\item Dipartimento di Fisica ``Enrico Fermi'', Universit\`a di Pisa, Pisa, 56127, Italy
\item INFN sezione di Pisa, Pisa, 56127, Italy
\item Dipartimento di Fisica, Universit\`a di Torino, Torino, 10125, Italy
\item INFN sezione di Torino, Torino, 10125, Italy
\item Institut des Hautes Etudes Scientifiques, 35 Route de Chartres, Bures-sur-Yvette, 91440, France
\item Niels Bohr International Academy, Niels Bohr Institute, Blegdamsvej 17, 2100 Copenhagen, Denmark
 \end{affiliations}

\begin{abstract}
  \textbf{
  Gravitational waves from $\sim 90$ black holes binary systems 
  have currently been detected by the LIGO~\cite{TheLIGOScientific:2014jea} and Virgo~\cite{TheVirgo:2014hva} experiments, 
  and their progenitors' properties inferred~\cite{LIGOScientific:2021djp}. 
  This allowed the scientific community to draw conclusions 
  on the formation channels of black holes in binaries, 
  informing population models and -- at times -- defying 
  our understanding of black hole astrophysics.
  The most challenging event detected so far is the short duration gravitational-wave transient GW190521~\cite{Abbott:2020tfl,Abbott:2020mjq}.
  We analyze this signal under the hypothesis that it was generated by the merger of two 
  nonspinning black holes on hyperbolic orbits.
  The best configuration matching the data corresponds to 
  two black holes of source frame masses of
  $81^{+62}_{-25}M_\odot$ and $52^{+32}_{-32}M_\odot$
  undergoing two encounters and then merging into an intermediate-mass  
  black hole.
  We find that the hyperbolic merger hypothesis is favored with respect to a quasi-circular
  merger with precessing spins with Bayes' factors larger than 4300 to 1, 
  although this number will be reduced by the currently uncertain prior odds.
  Our results suggest that GW190521 might be the first gravitational-wave detection 
  from the dynamical capture of two stellar-mass nonspinning black holes.}
\end{abstract}

\maketitle

\section{Introduction}
The gravitational-wave (GW) transient GW190521
is compatible with the quasi-circular merger of two heavy
($m_1 \simeq 85 \Msun$, $m_2 \simeq 66 \Msun$) black holes (BHs) resulting in an
${\simeq}150\Msun$ intermediate-mass BH (IMBH)~\cite{Abbott:2020tfl,Abbott:2020mjq}.
The estimated BH component masses fall in a mass gap ${\simeq} 65-120 \Msun$ 
for BHs formed directly from stellar collapse, and challenge
standard scenarios on BHs formation\cite{Abbott:2020mjq, LIGOScientific:2020kqk,Gonzalez:2020xah,Belczynski:2020bca,Mapelli:2021syv,Sedda:2021abh,Tagawa:2021ofj,DallAmico:2021umv}, suggesting the possibility of a progenitors formation through repeated mergers~\cite{2020ApJ...902L..26F,2021arXiv210704639F}.
The short duration ($\sim 0.1$~s) of GW190521 and the absence 
of a premerger signal, identified also by unmodeled (or weakly modeled) 
analyses\cite{TheLIGOScientific:2014jea}, are critical aspects for the
choice of waveform templates in matched filtering analyses and thus for the
interpretation of the source. For example, under the hypothesis of a
quasi-circular merger, matching the signal morphology requires fairly
large in-plane components of the individual BH spins and results in a
(weak) statistical evidence for orbital-plane precession.
High orbital eccentricities are also compatible with the 
burst-like morphology of GW190521, but best-matching eccentric 
merger waveforms still require spin precession\cite{Gayathri:2020coq, Romero-Shaw:2020thy}.
Spin precessing binary black hole (BBH) mergers are known to be degenerate with head-on collisions\cite{CalderonBustillo:2020odh}.
However, a head-on BBH is disfavored with respect to a boson-star head on collision with a log Bayes' factor of $-6.1$\cite{CalderonBustillo:2020srq}.
Other proposed interpretations involve a high-mass black hole-disk
system~\cite{Shibata:2021sau} or an intermediate mass ratio
inspiral\cite{Nitz:2020mga} (see also\cite{Estelles:2021jnz}),
and indicate that the origin of GW190521 is still unsettled.

In this \textit{Letter} we analyse GW190521 within the scenario of a
binary black hole (BBH) dynamical capture and 
compare this hypothesis to that of a quasi-circular merger.
Dynamical captures have a phenomenology radically different from
quasi-circular mergers\cite{East:2012xq,Gold:2012tk,Loutrel:2020kmm}.
The close passage and capture of the two objects in hyperbolic orbits naturally accounts 
for the short-duration, burst-like waveform morphology of GW190521 even in
the absence of spins.
Moreover, possibile explanations of the high component masses rely on
second-generation BHs, stellar mergers in young star clusters and BH
mergers in active galactic nuclei disks\cite{LIGOScientific:2020kqk,Rasskazov:2019gjw,Tagawa:2019osr,Gonzalez:2020xah,Belczynski:2020bca,Mapelli:2021syv,Sedda:2021abh,Tagawa:2021ofj,DallAmico:2021umv}, 
for which dynamical captures are possible. While no observational evidence for
GWs from dynamical captures existed prior to our work, such events are
not incompatible with the current detection rates~\cite{Rodriguez:2018pss,Mukherjee:2020hnm},
although these rates would require corrections to take into account the large masses of GW190521\cite{Mandel:2021smh}.

\section{Phenomenology of hyperbolic mergers}
A significant progress in constructing waveform templates for black hole binaries on 
hyperbolic orbits has been recently made within the effective-one-body (EOB) 
approach\cite{Chiaramello:2020ehz,Nagar:2020xsk}.
The EOB method\cite{Buonanno:1998gg}
 is a powerful analytical formalism that suitably resums post-Newtonian (PN) results\cite{Blanchet:2013haa,Schafer:2018kuf}
(obtained via a perturbative expansion of Einstein's field equations in powers of $v/c$, with $v$ the typical speed of the system) 
in the weak-field, small-velocity regime and makes them reliable and predictive also when the field is strong and velocities 
are comparable to $c$, i.e. up to merger and ringdown.
This framework can be extended to fully account for the dynamical capture phenomenology, and delivers complete waveform templates 
from hyperbolic mergers\cite{Chiaramello:2020ehz,Nagar:2020xsk}. 
The method is a generalization of the quasi-circular, spin-aligned waveform model \TEOBResumS{}\cite{Nagar:2020pcj} 
to deal with arbitrarily eccentric orbits,
from eccentric inspirals to hyperbolic mergers.
For simplicity, however, the EOB analytical waveform for hyperbolic
mergers does not contain  next-to-quasi-circular corrections informed by numerical relativity (NR)
simulations and it is completed by a NR-informed quasi-circular ringdown\cite{Nagar:2020pcj}.
The reason for this choice is that, although some NR simulations are 
available\cite{Pretorius:2007jn,Healy:2009zm,Sperhake:2009jz,East:2012xq,Gold:2012tk,Damour:2014afa,Hopper:2022rwo},
a systematic coverage of the BBH parameter space for hyperbolic orbits is currently missing. 

Although the model can be extended to include aligned spins and subdominant
multipoles in the waveforms, here we focus on nonspinning BBHs and use only the dominant
$\ell=m=2$ quadrupole mode, which has been more extensively tested , and was shown to be more than 97$\%$ faithful to NR (see Methods).
The inclusion of spins is expected to modify the most likely values of the parameters correlated with 
the spins, such as the mass ratio, but not e.g. the total mass of the binary.
The evidence of the analysis, too, is expected to vary due to the different prior volume explored. 
However, since the nonspinning model is contained within the spinning one, 
point estimates such as the maximum likelihood should not decrease with the addition of spin interactions.

The EOB relative motion is described using mass-reduced phase-space 
variables $(r,\varphi,p_\varphi,p_{r_*})$,  related to the physical
ones by [in geometric units $G=c=1$] $r=R/M$ (relative separation), $\varphi$ (orbital phase),
$p_{r_*}\equiv P_{R_*}/\mu$ (radial momentum), $p_\varphi\equiv P_\varphi/(\mu M)$
(angular momentum) and $t\equiv T/M$ (time), where $\mu\equiv m_1 m_2/M$ and $M\equiv m_1+m_2$.
The EOB Hamiltonian is $\hat{H}\equiv H/\mu\equiv \nu^{-1}\sqrt{1+2\nu(\hat{H}_{\rm eff}-1)}$, 
with $\nu\equiv \mu/M$ and $\hat{H}_{\rm eff}$ is the effective
Hamiltonian\cite{Nagar:2020pcj,Chiaramello:2020ehz,Nagar:2020xsk}.
For nonspinning binaries, the configuration space can be characterized by the 
mass ratio $q=m_1/m_2\geq 1$, the initial energy $E_0/M$ and the initial reduced 
orbital angular momentum $\pphi$~\cite{Nagar:2020xsk}.
Similarly to the motion of a test particle moving around a Schwarzschild BH, 
the EOB behavior of a hyperbolic encounter is characterized by the EOB potential
energy $E_{\rm EOB}\equiv M\sqrt{1+2\nu(\hat{W}_{\rm eff}-1)}$, where 
$\hat{W}_{\rm eff}=\sqrt{A(r)(1+p_\varphi^2/r^2)}$ is the effective potential energy.
Here, $A(r)\equiv P^1_5[A_{\rm 5PN}(r)]$  is the Pad\'e resummed EOB radial
potential, where $A_{\rm 5PN}(r)=1-2/r + \nu a(\nu,r)$  indicates its 
Taylor-expanded form, that reduces to the Schwarzschild case in the test-particle limit, $\nu=0$. 
The function $a(\nu,r)$ incorporates high-order corrections up to 5PN and it is 
additionally informed by NR simulations\cite{Nagar:2020pcj,Chiaramello:2020ehz}.
The solution $\de_r W_{\rm eff}=\de_r^2W_{\rm eff}=0$ defines last stable 
orbit (LSO) parameters $(r_{\rm LSO},p_\varphi^{\rm LSO})$.
When $p_\varphi >p_\varphi^{\rm LSO}$, $W_{\rm eff}$ has both a maximum and a minimum
and, depending on $E_0/M$, bound as well as unbound configurations are present.
In the absence of radiation reaction, unbound configurations are defined
by the condition $E_0/M>1$.
We define $E_{\rm min}/M \equiv \nu \hat{H}(r_0,q,p_\varphi, p_r=0)$ the energy corresponding 
to the initial separation and $E_{\rm max}/M = \max_r\left[\nu\hat{H}(r,q,p_\varphi, p_r=0)\right]$.
For a given $p_\varphi$, the values $(E_{\rm max},E_{\rm min})$ 
correspond respectively to unstable and stable circular orbits, analogously to
Schwarzschild geodesics.
When $E_0 > E_{\rm max}$ the objects fall directly onto each other without
forming metastable configurations (e.g., for head-on collisions, corresponding to $p_\varphi=0$).
When $1< E_0/M\leq E_{\rm max}/M$, the phenomenology changes from 
direct plunge, to on up to many close passages before merger, to
zoom-whirl behavior or even scattering\cite{Pretorius:2007jn,Sperhake:2009jz,Damour:2014afa}.

\begin{figure*}[h]
\centering
\includegraphics[angle=0,scale=0.5]{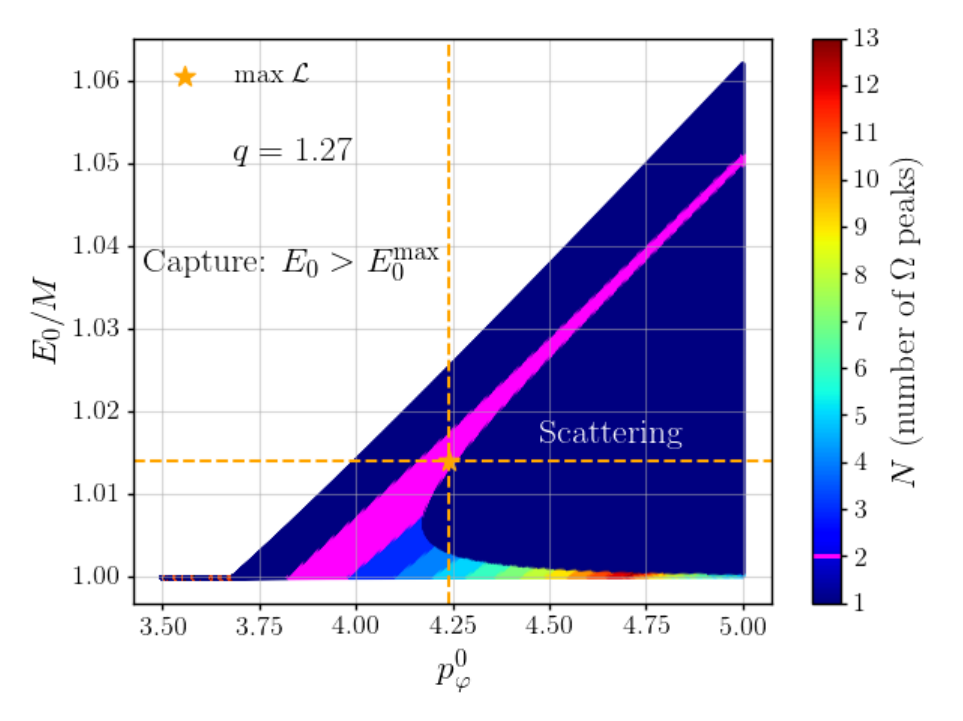}
\caption{{\bf Number of encounters as a function of the initial energy and angular momentum.} Parameter space for nonspinning hyperbolic encounters predicted using the \TEOBResumS{} 
EOB model and fixing $q \equiv m_1/m_2=1.27$. Here $(\pphi,E_0)$ are the EOB initial angular 
momentum and energy, while $E_{\rm max}^0$ is the value corresponding to unstable circular orbit. 
For $E_0<E_0^{\rm max}$, each color labels the number of peaks (i.e. of periastron passages) $N$ 
of the EOB orbital frequency $\Omega$. The orange star labels the maximum likelihood values 
$(\bar{p}_{\varphi}^0, \bar{E}_0)$ corresponding to the {\it constrained} analysis, see Table~\ref{tab:gw190521}.}
\label{fig:EJ_parspace}
\end{figure*}

In the presence of radiation reaction, the qualitative picture remains unchanged 
(as also observed in NR simulations\cite{Gold:2012tk}), although the threshold 
between the two qualitative behaviors is not simply set by  
$E_{\rm max}$, but it is also affected by GW losses.
The latter are taken into account through the azimuthal and radial radiation
reaction forces $({\cal F} _\varphi,{\cal F}_r)$ described
in detail in\cite{Chiaramello:2020ehz,Nagar:2020xsk}. 
The dynamics of each configuration can be characterized by counting 
the number of peaks of the orbital frequency
$\Omega(t)\equiv \dot{\varphi}$, each peak corresponding to a
periastron passage\cite{Nagar:2020xsk}.
Figure~\ref{fig:EJ_parspace} illustrates the $(E_0/M,\pphi)$ parameter space, 
defined using the peaks of $\Omega(t)$, of a nonspinning binary with $q=1.27$, corresponding 
to the best-matching mass ratio for the GW190521 analysis.
The different colors indicate the number of encounters. 
Although the two dark blue areas, above and below the magenta zone, possess a single 
peak in $\Omega$, they correspond to different phenomenologies. The dark-blue region {\it above}
the magenta area corresponds to a direct capture scenario, that eventually 
leads to a ringdown phase. The dark-blue region {\it below} the magenta area 
corresponds to a scattering scenario. 
The single-burst waveform morphology  is obtained for $(E_0/M,\pphi)$ within the blue capture region 
as well as the upper boundary of the magenta region, until a distinct second burst
of GWs does not appear before the one corresponding to the final merger.
The single burst phenomenology also occurs in the white region $E_0 > E_{\rm max}$
in Fig.~\ref{fig:EJ_parspace}, where there exist systems with low 
values of $\pphi$ and large  initial energies. Waveforms emitted by such binaries 
are dominated by the ringdown.



\section{Results}
\begin{figure*}[h]
\begin{center}
\includegraphics[width=0.5\textwidth]{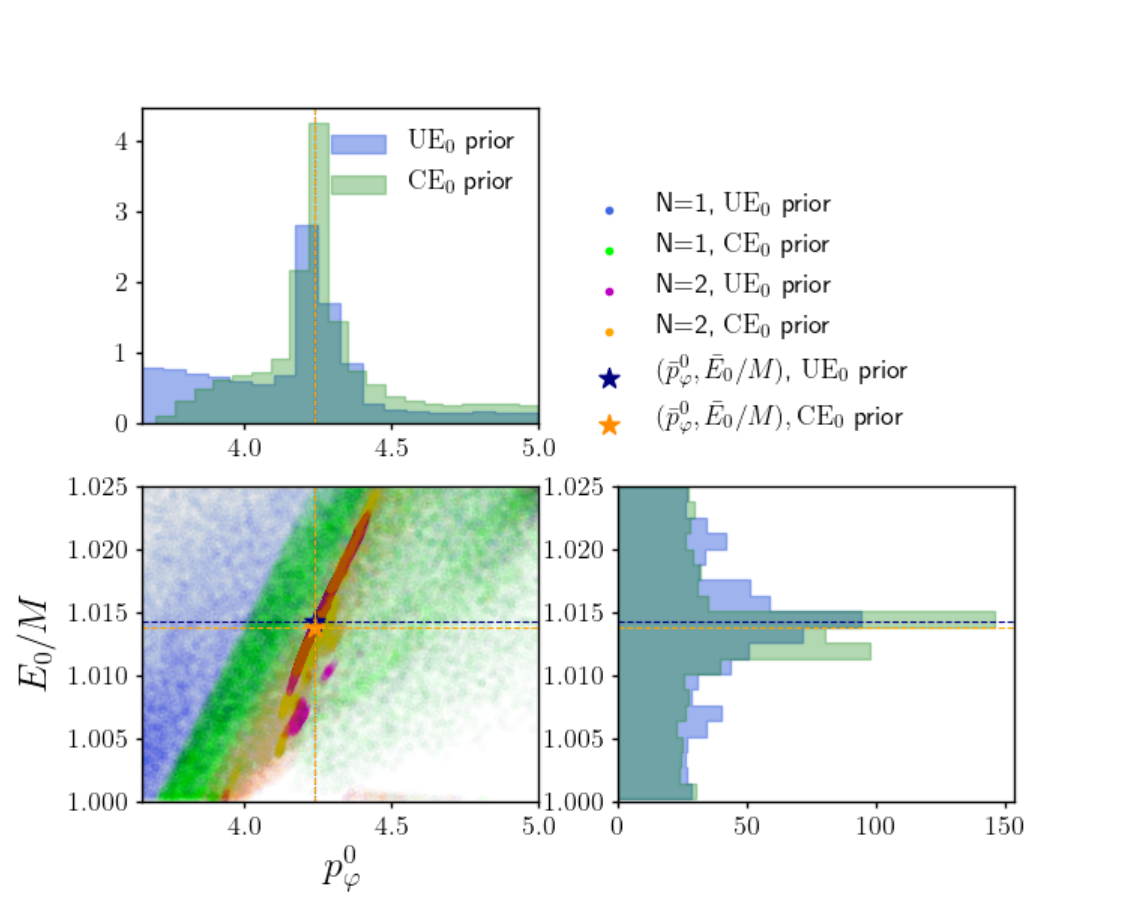}
\caption{ {\bf Energy-momentum marginalized two dimensional posterior.} Marginalized two-dimensional posterior distributions of the initial energy
	$E_0$ and initial angular momentum $\pphi$ for the constrained ($\CE$) and
	unconstrained ($\UE$) energy prior choices.
	The colors highlight the different waveform phenomenologies, 
	with $N=1$ (blue and green) or $N=2$ (magenta and orange) peaks in the orbital frequency.
	The maximum likelihood values $(\bar{p}_{\varphi}^0, \bar{E}_0)$ are highlighted with red 
	($\UE$ prior) and dark-orange ($\CE$ prior) stars.}
\label{fig:EJ_gw190521}
\end{center}
\end{figure*}

We analyze GW190521 under the hypothesis that it was generated by a dynamical 
capture of two nonspinning BHs. We use two different priors for the initial energy
of the binary: an ``unconstrained" prior (UE$_0$) and a ``constrained" prior CE$_0$ (see Methods).
The results of the analysis corresponding to the UE$_0$ and CE$_0$
priors are summarized in the second 
and third columns of Table~\ref{tab:gw190521} respectively.
The consistency of the two measurements confirms 
the robustness of our modeling choices. Focusing on global fitting quantities 
we find, respectively for the UE$_0$ (CE$_0$) priors, maximum likelihood values $\log({\cal L})_{\rm max} = 123.2 \, (123.0)$, 
and Bayesian evidences $\log{\cal B}^{\rm signal}_{\rm noise} = 84.0 \pm 0.18 \, (83.3 \pm 0.18)$, 
while the recovered matched-filter signal-to-noise ratio (SNR) is equal to 15.2 (15.4).
Employing the standard cosmology~\cite{Aghanim:2018eyx}, we find component masses in the 
source frame $(m_1,m_2) = (85^{+88}_{-22},59^{+18}_{-37})\Msun$
for the UE$_0$ case and $(m_1,m_2)=(81^{+62}_{- 25},52^{+32}_{-32})M_\odot$
in the CE$_0$ case.
Figure~\ref{fig:EJ_gw190521} illustrates the $(E_0,\pphi)$ parameter space selected
by the analysis, with colors highlighting configurations with different number of encounters $N$. The figure shows 
that, despite GW190521 consisting of a single GW burst around the
analyzed time, many of the configurations selected, and in particular the most probable ones, correspond 
to two encounters.

\begin{figure*}[h]
\begin{center}
\includegraphics[width=0.44\textwidth]{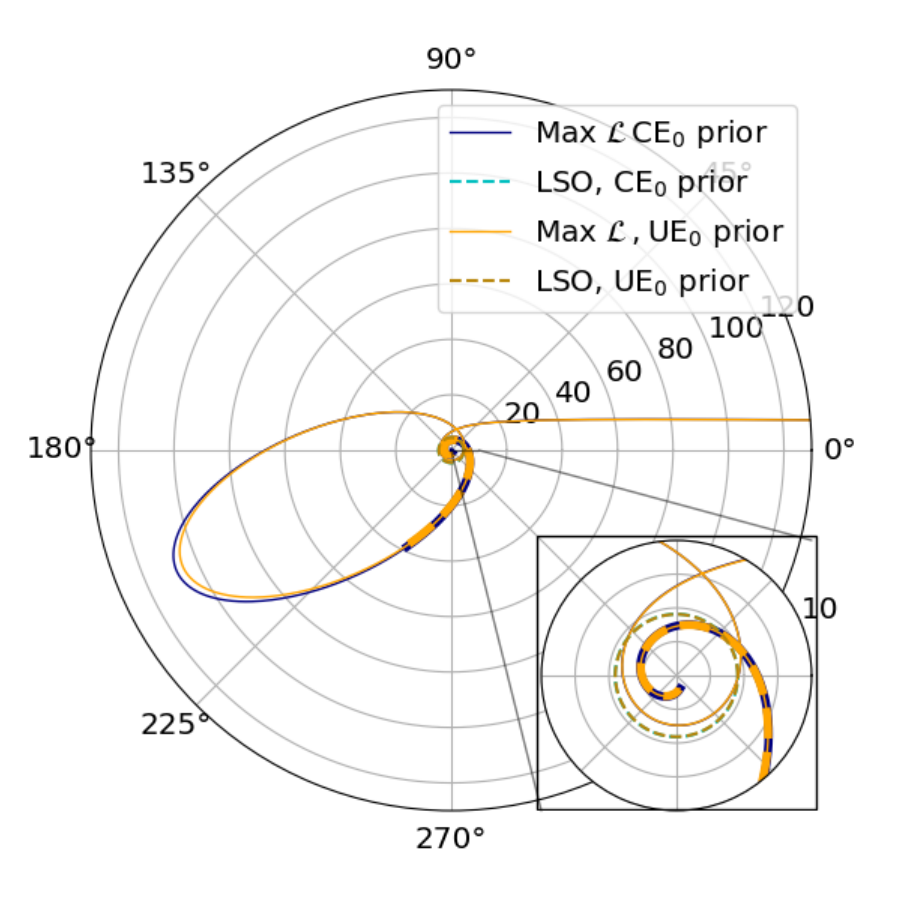}
\includegraphics[width=0.54\textwidth]{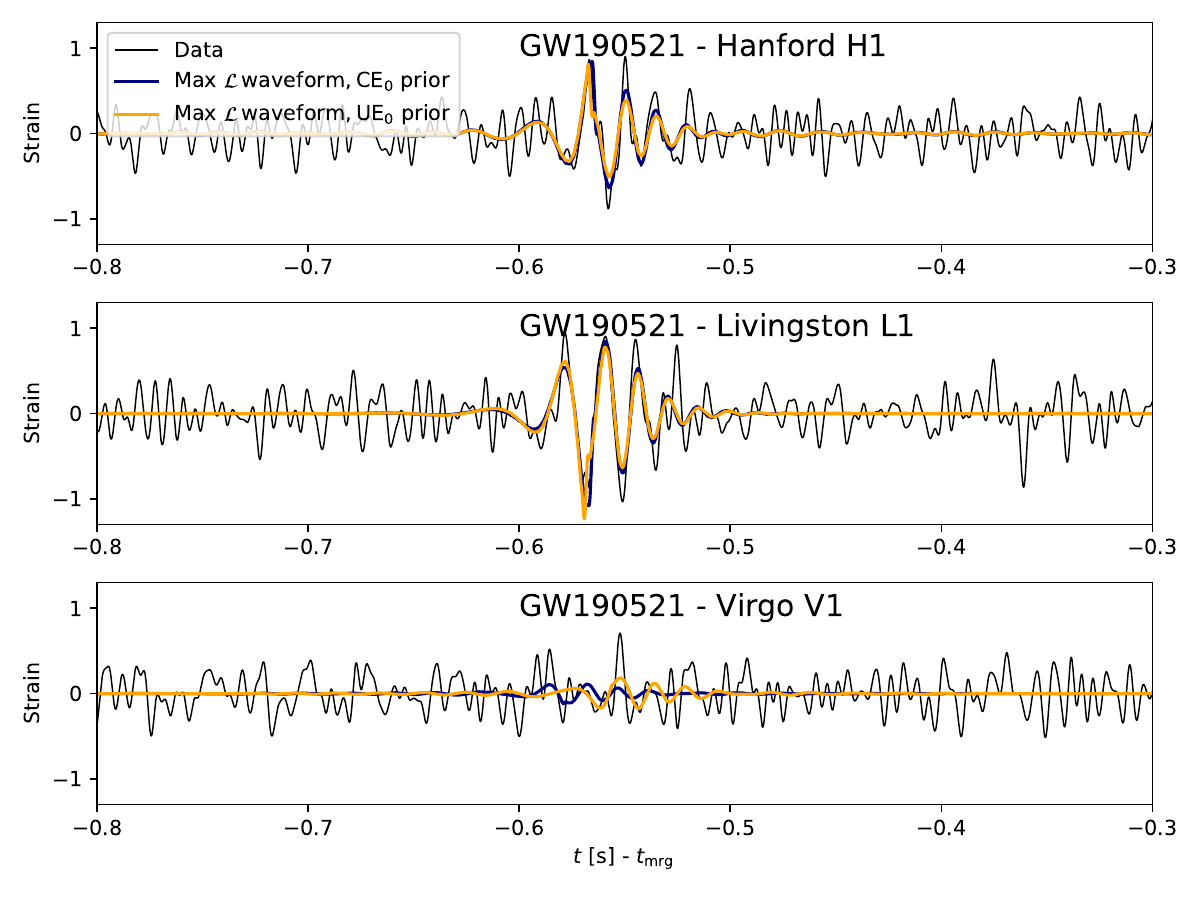}
\caption{\textbf{Maximum likelihood configurations with the two different energy priors,
$\UE$ (orange) and $\CE$ (blue)}. Top: the  $(r,\varphi)$ EOB relative orbit. 
Bottom: the waveform templates projected onto the three detectors compared 
to the whitened LIGO-Virgo data around the time of GW190521. 
The most probable last stable orbits (LSO) are highlighted with gold ($\UE$ prior) and cyan ($\CE$ prior) dashed lines 
and are located, respectively, at $\bar{r}_{\rm LSO}= 4.54$ and $\bar{r}_{\rm LSO}=4.52$. 
Corresponding mass ratios are $\bar{q}=1.04$ and $\bar{q}=1.27$.
The inset highlights the first close encounter, that is then followed by a highly eccentric orbit that eventually 
ends up with a plunge and merger phase. The part of the trajectory from 
$\sim (t_{\rm GPS} - 0.8$~s$)$ to merger, which contributes to the second GW burst, 
is highlighted with thicker lines in the plot.
Note that the GW bursts corresponding to the first encounter occur ${\sim}4$~s before 
the GW190521 time, their magnitude is comparable to the detector noise 
and are outside the segment of data analyzed.}
\label{fig:gw190521}
\end{center}
\end{figure*}

The phenomenology corresponding to the set of maximum likelihood parameters 
selected by the analysis are shown in Fig.~\ref{fig:gw190521}. The EOB relative trajectory (top panel)
is complemented by the corresponding waveform templates projected onto the three detectors and 
compared to the whitened LIGO-Virgo data around the time of GW190521.
Thicker lines highlight the last part of the dynamics, which exactly covers 
the portion of the signal displayed in the bottom panel.
The magnitude of the first GW burst predicted by the EOB analysis 
(not shown in the plot) is comparable to the detector noise and would
occur outside the analysis window. However, we find that such 
first burst is not a robust feature across samples, occurring 
at different times and smaller amplitudes for different points and
not occuring at all for others (see Fig.~\ref{fig:EJ_gw190521}).
Given this consideration and the small amplitude of such first burst, we
do not expect an extension of the analysis segment to impact our main conclusions.

  In order to compare the hyperbolic capture with the quasi-circular
  merger hypothesis, we perform a new quasi-circular analysis with the precessing surrogate 
model {\tt NRSur7dq4}\cite{Varma:2019csw} and with the
  quasi-circular precessing flavor of \TEOBResumS{}, \TEOBResumSP{}
  \cite{Akcay:2020qrj, Gamba:2021ydi}.
  To minimize systematic effects,
  we consistently use the {\tt bajes} pipeline\cite{Breschi:2021wzr}
  with the same settings discussed above for all the runs.
The prior distributions for the mass parameters and the extrinsic
parameters are also identical to the ones used in the hyperbolic capture
analysis with {\tt TEOBResumS}, while the prior
on the spin components is chosen to be uniform in the spin magnitudes and isotropic in the angles\cite{Abbott:2020tfl}.
When including higher modes we disable phase marginalization. 

In Table~\ref{tab:gw190521} we quote maximum likelihood and
matched-filter SNR values obtained from the full unmarginalised posterior.
The quasi-circular precessing analyses with {\tt bajes} and {\tt NRSur7dq4} 
are in agreement with those obtained by LVK\cite{Abbott:2020tfl}, confirming the reliability of the infrastructure adopted for the 
inference. The maximum SNR recovered via our pipeline is lower by $0.7$
than the one extracted from the public LVK samples. We attribute this discrepancy to 
differences in data-processing between pipelines, small differences in the prior boundaries and the sampling itself.
The use of consistent settings in our new runs with the model used by the LVK
excludes that such discrepancies affect the comparison against the non-circular analysis.

The {\tt TEOBResumSP} analyses display consistency with the NR surrogate. 
When PE is performed with the dominant $(2,2)$ mode, it yields $\log\mathcal{L}_{\rm max} = 106.0$, 
$\rm{SNR}_{\rm max} = 14.7$ and $\log\mathcal{B}^{\rm signal}_{\rm noise} = 72.95 \pm 0.08$. 
This indicates that the observed increases in these statistics when employing the dynamical capture model 
are not driven by subtle differences between waveform families (EOB and NR surrogate).

\section{Discussion}

Despite the different hypotheses on the coalescence process, 
our results on the component masses are in good agreement with the ones 
obtained from a quasi-circular model.
This confirms that an IMBH is formed at the end of the coalescence
also in the hyperbolic merger scenario.
The consistency on the total mass is not surprising, given that the 
dominant contribution to this parameter comes from the determination of
the ringdown frequency~\cite{Abbott:2020tfl}. However, the dynamical capture
model is able to fit GW data better than the quasi-circular scenario despite having four less degrees of freedom, 
with a 16 e-fold increase in the maximum likelihood value.
For comparison, the distribution of $\log\mathcal{L}$ of the quasi-circular analysis
spans a $\sim 26$ e-folds range and has median $\sim 9$ e-folds smaller than its maximum
value. If we assume this difference to be representative of the statistical uncertainty 
$\sigma$ on the likelihood, we find that our result lies about $3 \sigma$ from the median
of the quasi-circular analysis and $2 \sigma$ from its maximum value.
Under the dynamical capture assumption, we obtain a matched-filter SNR $\rho=15.4$,
larger by almost a unity with respect to the same value obtained using quasi-circular waveforms. Similarly to the $\log\mathcal{L}$, the maximum SNR of the hyperbolic analysis lies about $1 \sigma$ from the corresponding value of the quasi-circular analysis and $2 \sigma$ from its median.

The fit improvement registered by these two indicators is confirmed by the Bayesian evidences, 
keeping into account the full correlation structure of the parameter
space, which imply odds ${\gtrsim}4315{:}1$ in favor of the dynamical
encounter scenario against the quasi-circular scenario.
This number is expected to be an optimistic estimate of the posterior odds, due to the prior odds
disfavouring a dynamical capture scenario compared to a quasi-circular binary. 
However, estimates of prior odds are currently not reliable due to
orders of magnitudes uncertainties on dynamical capture rates in this
mass range~\cite{Mandel:2021smh} and, as such, we do not attempt to
quantify them \red{directly}.
Given our (conservative) Bayes factor, we estimate that the capture interpretation
is favored with respect to a quasi-circular stellar-collapse scenario\cite{LIGOScientific:2020kqk} so long as the rates of 
such events is larger than $5 \times 10^{-3} \rm{Gpc}^{-3} \rm{yr}^{-1}$. This number is computed by imposing that the posterior odds are lager 
than one, i.e. that $4315 \times R^{\rm dc}/R^{\rm qc} > 1$, where $R^{\rm dc}$ is the rate of dynamical capture events and 
$R^{\rm qc}= 23.9$ Gpc${}^{-3}$ yr${}^{-1}$ as estimated by LVK\cite{LIGOScientific:2020kqk}.
Notably, the Bayes' factors
receive a penalty disfavouring the quasi-circular hypothesis
due to the larger dimensionality of this model which is not phase marginalized and includes precessing spins degrees of freedom,
although the latter are only weakly measurable.
Additionally, some railing against the prior can be observed for the $E_0/M$ and the UE$_0$ $p_\varphi^0$
posterior samples, which might affect the estimation of the dynamical capture evidence.
However, the choice of prior bounds in this analysis was dictated either by physical boundaries, and hence cannot be relaxed,
or by considerations on computational cost and model validity in a region 
-- that of head-on mergers -- which was shown to have 
little support for the phenomenology observed\cite{CalderonBustillo:2020srq}.
In light of the above caveats, the Bayesian evidences alone represent useful, but not decisive 
proof in favor of the capture scenario.

Nonetheless, these two results combined constitute data-driven 
indicators that the interpretation of GW190521 within the dynamical capture 
scenario seems preferred over a quasi-circular spin-precessing 
merger\cite{TheLIGOScientific:2014jea,Estelles:2021jnz}.
No other analysis shows such large improvements in evidence and log-likelihood with respect to the equal-mass, 
quasi-circular scenario\cite{Gayathri:2020coq,Romero-Shaw:2020thy,CalderonBustillo:2020srq}.
At the same time, the \textit{absolute} values of evidence and maximum likelihood estimated 
in some studies\cite{Estelles:2021jnz} are almost as large as those obtained in this work. 
These values were however obtained with a model which is less NR-faithful in the quasi-circular case than the 
{\tt NRSur7dq4} model considered in this work. Although a direct comparison is not possible given the different 
PE infrastructure, sampler, models and priors explored, this fact highlights
the necessity of exploring multiple hypothesis and model selection
to understand such short GW transients.
Our findings are consistent with the fact that burst-like waveforms from
highly eccentric or head-on BBH collision may be confused with mildly
precessing quasi-circular binaries\cite{CalderonBustillo:2020odh} and viceversa. 
In the supplementary material we confirm this degeneracy to a certain extent, but we
show that the preference we obtain for the non-circular model is incompatible with the true signal
being a quasi-circular merger embedded in gaussian noise.
Regarding other possible scenarios, a quantitative comparison is
currently not possible since they 
have not been analyzed with full Bayesian
studies and/or complete waveform 
templates\cite{CalderonBustillo:2020srq,Gayathri:2020coq,Shibata:2021sau}.

While our analysis selects a two-encounters merger as best-fitting
capture scenario (Fig.~\ref{fig:gw190521}), the orbital dynamics of these
encounters is rather sensitive to changes in both the conservative and
nonconservative part of the dynamics\cite{Nagar:2020xsk}, as also evident
from Fig.~11 of\cite{Nagar:2020xsk}.
Going beyond the conservative assumptions behind our analysis,
future work will explore the impact of spin and of higher waveform multipoles, 
as well as consider systematic comparisons between our (improved\cite{Nagar:2021gss}) 
EOB model and a larger number of NR simulations. The inclusion of additional, physically motivated, degrees 
of freedom (e.g., BH spins) is expected to further shed light on the nature of GW190521.

\section{Methods}

\subsection{Waveform model validation}

The EOB analytical model employed in this analysis, {\tt TEOBResumS} \cite{Chiaramello:2020ehz,Nagar:2020xsk} , generates waveforms and
scattering angles that are faithful to NR simulations of nonspinning BBH along eccentric and
scattering orbits\cite{Damour:2014afa,Chiaramello:2020ehz,Nagar:2020xsk}.
The model is directly validated in the regime of interest by comparisons against new NR data targeted at GW190521.
The simulations parameters are chosen to be compatible with the ones obtained by our hyperbolic analysis.
Additionally, we compare against 46 selected nonspinning, highly eccentric NR simulations\cite{Healy:2022wdn},
to validate the model in a similar regime.
Crucially, for all configurations considered,
the quasi-circular ringdown provides a reliable approximation of the
final stage of the coalescence, and the model is more than $97\%$ faithful to NR.
This is not surprising, and can be attributed to the circularization of the system during the last phases of the coalescence. 
Such results ensure the reliability of our model in extracting astrophysical properties from GW signals.
Head-on collisions, conversely, are not well approximated by our model.
Finally, reliability of this EOB approach in describing dynamical captures
is further verified in the test-mass limit of a body
captured by a Schwarzschild BH, using the waveforms computed numerically using black hole
perturbation theory\cite{Harms:2014dqa,Albanesi:2021rby}.
These results are summarized in the Supplementary Material; 
further details and more in-depth comparisons will also be presented in a future work (Andrade et. al., in preparation).

\subsection{GW190521 analysis}
The publicly released GW190521 data are analyzed around time
$t_{\rm GPS} = 1242442968$, with an $8\,$s time-window and in the range of
frequencies $[11,512]\,$Hz using the {\tt bajes}
pipeline~\cite{Breschi:2021wzr}.
We employ the power-spectral-density estimate and calibration envelopes 
publicly available from the GW Open Science
Center~\cite{Abbott:2019ebz}.
The Bayesian analysis uses the {\tt dynesty}
sampler~\cite{Speagle:2020} with 2048 live points. 
We use a uniform prior in the mass components $(m_1,m_2)$ exploring the ranges of chirp mass 
$\mathcal{M}_c \in [30, 200]\,M_{\odot}$ and mass ratio $q \in [1, 8]$.
The luminosity distance is sampled assuming a volumetric prior in the range $[1, 10]\,$Gpc.
We analytically marginalize over the coalescence phase, and sample the coalescence
time in $t_s \in [-2,2]\,$s with respect to the central GPS time.

The key quantities to sample the configuration space of hyperbolic mergers
are  $(E_0/M,\pphi)$.
The initial angular momentum is uniformly sampled within $\pphi\in[3.5, 5]$, and
further imposing $\pphi \geq p_{\varphi}^{\rm LSO}$ for any $q$.
The initial energy is uniformly sampled in the interval
$E_0/M \in [1.0002, 1.025]$ but with two different additional
constraints that result in two different prior choices:
($\UE$) \textit{Unconstrained prior}: $E_0\geq E_{\rm min}$;
($\CE$) \textit{Constrained prior}: $E_{\rm min}\leq E_0\leq E_{\rm max}$. 
The $\UE$ prior spans a larger portion of the parameter space, notably including
direct capture, although the dynamic remains far from the head-on
collision case. The $\CE$ prior is contained in the first, and restricts the parameter space to
systems closer to stable configurations, for which the orbital dynamics substantially contributes 
to the waveform and the ringdown description is expected to be more accurate.

\noindent{\bf Data Availability} 
\noindent Data is available onn Zenodo, with DOI {\tt 10.5281/zenodo.7081337}.

\noindent{\bf Code Availability} 
\noindent The eccentric waveform model used in this work, {\tt TEOBResumS}, 
 is publicly available at: \url{https://bitbucket.org/eob_ihes/teobresums/} 
 and results presented in this paper have been obtained with the
 version tagged ${\tt eccentric.v0\_a6c\_c3\_circularized}$. 
 Similarly, {\tt TEOBResumSP} is publicly availabe at the same address, and results presented
 here have been obtained with the version having git hash ${\tt 56f20ad}$. 

\noindent{\bf Acknowledgments}
 We are grateful to T.~Damour, J.~A.~Font and T.~Andrade for discussions.
 We are also grateful to D.~Chiaramello for collaboration at the beginning of the 
 project. We also thank B.~Daszuta, F.~Zappa, W.~Cook and D.~Radice for
 supporting the development of the {\tt GR-Athena++} code and for help with
 the NR simulations presented in the Supplementary Material.
 R.G. acknowledges support from the Deutsche Forschungsgemeinschaft
 (DFG) under Grant No. 406116891 within the Research Training Group
 RTG 2522/1. 
 M.B., S.B. acknowledge support by the EU H2020 under ERC Starting
 Grant, no.~BinGraSp-714626.
 M.B. acknowledges partial support from the Deutsche Forschungsgemeinschaft
 (DFG) under Grant No. 406116891 within the Research Training Group
 RTG 2522/1.
 G.C. acknowledges support by the Della Riccia Foundation under an Early Career Scientist
 Fellowship, and funding from the European Union’s Horizon 2020 research and innovation program
 under the Marie Sklodowska-Curie grant agreement No. 847523 ‘INTERACTIONS’, from the Villum Investigator program supported by VILLUM FONDEN (grant no. 37766) and the DNRF Chair, by the
 Danish Research Foundation.
    Computations were performed on the national HPE
 Apollo Hawk at the High Performance Computing Center Stuttgart (HLRS), on the ARA cluster at Friedrich
 Schiller University Jena and on the {\tt Tullio} sever at INFN Turin. The ARA cluster is funded in
 part by DFG grants INST 275/334-1 FUGG and INST275/363-1 FUGG and by the ERC Starting Grant, 
 grant agreement
 no. BinGraSp-714626. The authors acknowledge
 HLRS for funding this project by providing access to the supercomputer HPE Apollo Hawk under the grant
 number INTRHYGUE/44215.
 We thank E.~Ferrari for speed-up coding work on {\tt Tullio}.\\
 This research has made use of data, software and/or web tools obtained 
 from the Gravitational Wave Open Science Center (\url{https://www.gw-openscience.org}), 
 a service of LIGO Laboratory, the LIGO Scientific Collaboration and the 
 Virgo Collaboration. LIGO is funded by the U.S. National Science Foundation. 
 Virgo is funded by the French Centre National de Recherche Scientifique (CNRS), 
 the Italian Istituto Nazionale della Fisica Nucleare (INFN) and the 
 Dutch Nikhef, with contributions by Polish and Hungarian institutes.

\noindent{\bf Author contributions}
\noindent S.B and A.N. contributed to the origination of the idea. A.N., P.R., R.G., and S.B. developed and tested the waveform model. R.G., M.B., and G.C. performed the analyses and 
S.A. carried out Numerical Relativity simulations. R.G. produced all the figures. All authors worked out collaboratively the general details of the project. All authors helped edit the manuscript.

\noindent{\bf Competing interests}
The authors declare no competing interests.

\newpage     

\begin{table*}[h!]
\resizebox{\textwidth}{!}{\begin{tabular}{lccccc|ccc}
\hline
\hline
Reference & \multicolumn{5}{c|}{This paper} & LVK\cite{Abbott:2020tfl} & Gayathri et al.\cite{Gayathri:2020coq} & Romero-Shaw et al.\cite{Romero-Shaw:2020thy}  \\
\hline

Waveform                           & \TEOBResumS{}\cite{Chiaramello:2020ehz,Nagar:2020xsk}&  \TEOBResumS{}\cite{Chiaramello:2020ehz,Nagar:2020xsk} & {\tt TEOBResumSP}\cite{Gamba:2021ydi}\footnote{Spin results obtained at a reference frequency of 5 Hz.}  & {\tt NRSur7dq4}\cite{Varma:2019csw}& {\tt NRSur7dq4}\cite{Varma:2019csw}  & {\tt NRSur7dq4}\cite{Varma:2019csw} & {\tt NR}\cite{Healy:2022wdn} & {\tt SEOBNRE}\cite{Cao:2017ndf} \\
$E_0$ prior & 
Unconstrained ($\UE$) & 
Constrained  ($\CE$) & --& --&--&--  & --  \\
Multipoles & $(\ell,|m|)=(2,2)$ & $(\ell,|m|)=(2,2)$ & $(\ell,|m|)=(2,2)$ & $(\ell,|m|)=(2,2)$ & $\ell \le 4$ & $\ell\le4$ & --  & --  \\
\hline
$m_1$ [$\Msun$]              &  $85^{+88}_{-22}$          &  $81^{+62}_{-25}$      & $90^{+19}_{-14}$        & $102^{+35}_{-23}$       & $84^{+17}_{-12}$         &  $85^{+21}_{-14}$      & $102^{+7}_{-11}$   & $92_{-16}^{+26}$   \\
$m_2$ [$\Msun$]              & $59^{+18}_{-37}$           &  $52^{+32}_{-32}$      & $66^{+10}_{-8}$         & $64^{+19}_{-25}$        & $71^{+16}_{-18}$         &  $66^{+17}_{-18}$      & $102^{+7}_{-11}$   & $69_{-19}^{+18}$   \\
$M_{\rm source}$ [$\Msun$]   & $151^{+73}_{-51}$          &  $130^{+75}_{-43}$     & $156^{+25}_{-15}$       & $164^{+40}_{-23}$       & $153^{+29}_{-19}$        &  $150^{+29}_{-17}$     &  --                & --  \\ 
$m_2/m_1\leq 1$              & $0.69^{+0.27}_{-0.52}$     & $0.63^{+0.31}_{-0.43}$ & $0.73^{+0.21}_{-0.15}$  & $0.62^{+0.32}_{-0.30}$  & $0.86^{+0.12}_{- 0.30}$  & $0.79^{+0.19}_{-0.29}$ & --                 & -- \\
$\chi_{\rm eff}$             &  --                        &                     -- & $-0.05^{+0.09}_{-0.12}$ & $0.01^{+0.24}_{-0.26}$  & $-0.03^{+ 0.25}_{-0.26}$ & $0.08^{+0.27}_{-0.36}$ & $0$                &  $ 0.0_{-0.2}^{+0.2}$ \\
$\chi_{\rm p}$               &  --                        &                     -- & $0.72^{+0.16}_{-0.22}$  & $0.71^{+ 0.22}_{-0.36}$ & $0.79^{+0.16}_{-0.40}$   & $0.68^{+0.25}_{-0.37}$ & 0.7                & --  \\
$e$                          & --                         & --                     & --     & --                      & --                       & --                     & $0.67$             & $0.11$\footnote{Lower limit at 10~Hz.}  \\
$E^0/M$                      & $1.014^{+ 0.009}_{-0.012}$ &$1.014^{+0.010}_{-0.012}$ &-- &-- &--  &-- &-- & --  \\
$\pphi$                      & $4.18^{+0.50}_{-0.62}$     &$4.24^{+ 0.57}_{-0.37}$ & -- & --   &-- &-- &-- & -- \\
$D_L$ [Gpc]                  & $4.7^{+4.8}_{-2.7}$        &  $6.1^{+ 3.3}_{-3.7}$  & $4.5^{+1.2}_{-1.2}$     & $3.9^{+2.3}_{-1.9}$     & $4.8^{+2.3}_{-2.2}$      & $5.3^{+2.4}_{-2.6}$    & $1.84^{+1.07}_{-0.054}$ & $4.1^{+1.8}_{-1.8}$ \\
SNR${}_{\rm max}$            &  $ 15.2$                   & $15.4$                 & $14.7$                  & $14.7$                  &  $14.6$                  & $15.4$                 &   --                    &  -- \\
$\log({\cal L})_{\rm max}$   &  $123.2$                   & $123.0$                & $106.0$                 &  $107.0$                &  $105.6$                 &    --                  &   --                    &  -- \\
$\log{\cal B}^{\rm signal}_{\rm noise}$  &$84.00 \pm 0.18$&   $83.30 \pm 0.18 $    & $72.95 \pm 0.08$        & $74.76 \pm 0.11$        &  $74.86 \pm 0.11$        &   --                   &   --                    &  -- \\
\hline
\hline
\end{tabular}}
\caption{{\bf Source parameters of GW190521.} We indicate the mass of the heavier (lighter) object  with $m_1$ ($m_2$), 
$M_{\rm source}$ is the total mass in the frame of the source,
$\chi_{\rm eff}$ is the effective spin along the orbital angular momentum, 
while $\chi_{\rm p}$ is the effective precessing spin\cite{Abbott:2020tfl}.
The second and third columns report our new results,
obtained with the hyperbolic capture model\cite{Nagar:2020xsk} with the two different
prior choices on the energy. The fourth, fifth and sixth columns report the results obtained in this work 
with the quasi-circular models {\tt TEOBResumSP} and  {\tt NRSur7dq4}. 
For reference, the remaining columns report results of other analyses\cite{Abbott:2020tfl,Gayathri:2020coq, Romero-Shaw:2020thy}.
We employ the standard cosmology of Planck\cite{Aghanim:2018eyx} to compute source frame masses. 
Median values and $90\%$ credible intervals are quoted and natural
logarithms are reported. The SNR values correspond to the matched-filter estimates.}
\label{tab:gw190521}
\end{table*}     

\if\IncludeSM1

\clearpage

\setcounter{equation}{0}
\setcounter{figure}{0}
\setcounter{table}{0}
\setcounter{page}{1}
\makeatletter
\renewcommand{\theequation}{S\arabic{equation}}
\renewcommand{\figurename}{SUPPLEMENTARY FIGURE}
\renewcommand{\tablename}{SUPPLEMENTARY TABLE}

\begin{center}
\textbf{\large Supplemental Material}
\end{center}
\ifx\IncludeSM\undefined
\def\IncludeSM{0}
\fi

\if\IncludeSM0 \documentclass[aps,prl,onecolumn,nofootinbib,floatfix]{revtex4}
\usepackage{color}
\usepackage{amssymb}
\usepackage{acronym}
\usepackage{amsmath}
\usepackage{hyperref}
\usepackage{graphicx}
\usepackage{lineno}
\setcitestyle{super}
\usepackage{bm}
\usepackage{multibib}
\newcites{supp}{Supplementary References}

\hypersetup{
    pdfnewwindow=true,      
    colorlinks=true,       
    linkcolor=blue,          
    citecolor=blue,        
    filecolor=blue,      
    urlcolor=blue           
}


\allowdisplaybreaks

\def\p{\partial}
\def\Lie{{\cal L}}
\def\Lam{\Lambda}
\def\blam{\bar{\lambda}}
\def\Lamt{\tilde{\Lambda}}
\def\dLamt{\tilde{\delta{\Lambda}}}
\def\non{\nonumber}                     
\def\half{\frac{1}{2}}
\def\e{{\rm e}}
\def\i{{\rm i}}
\def\gccm{g\,cm^{-3}}
\def\Msun{M_{\odot}}
\def\GMc2{G M_{\odot} c^{-2}}
\def\Mpc{Mpc}
\def\eps{\epsilon}
\def\sf{{\rm SF}}
\def\I{\mathcal{I}}
\def\M{\mathcal{M}}
\def\O{\mathcal{O}}
\def\vareps{\varepsilon}
\def\vrho{\varrho}
\def\rad{rad}
\def\lm{{\ell m}}
\def\teobLR{{t^{\rm EOB}_{\Omega \, \rm peak}}}
\def\teobpo{{t^{\rm EOB}_{\Omega_{\rm orb}^{\rm max}}}}
\def\tdomgmax{{t^{\rm NR}_{\dot{\omega}_{22}\, \rm peak}}}
\def\tAmax{{t_{A_{22}^{\rm max}}^{\rm NR}}}
\def\tnrextr{{t_{\rm extr}^{\rm NR}}}
\def\teobNQC{{t^{\rm EOB}_{\rm NQC}}}
\def\lm{{\ell m}}
\def\v{v_\varphi}
\def\non{\nonumber}                  
\def\half{\frac{1}{2}}
\def\de{\partial}
\def\lm{{\ell m}}
\def\g{{\gamma}}
\def\o{{\rm o}}
\def\ii{{\rm i}}
\def\l{{\ell }}
\def\r{{\hat{r}}}
\def\ph{\varphi}
\def\th{\vartheta}
\def\A{{\cal A}}
\def\I{{\cal I}}
\def\F{{\cal F}}
\def\J{{\cal J}}
\def\U{{\cal U}}
\def\E{{\cal E}}
\def\B{{\cal B}}
\def\M{{\cal M}}
\def\S{{\cal S}}
\def\O{{\cal O}}
\def\X{{\bf X}}
\def\Y{{\cal Y}}
\def\N{{\cal N}}
\def\W{{\cal W}}
\def\Z{{\cal Z}}
\def\R{{\cal R}}
\def\k{{\hat{\hat{k}}}}
\def\a{\alpha}
\def\b{\beta}
\newcommand\be{\begin{equation}}
\newcommand\ee{\end{equation}}
\def\hr{\hat{r}}
\def\ha{{\hat{a}}}
\def\rmd{{\rm d}}
\def\rcm{R_*}
\def\rhoo{S_\pm}
\def\rhoi{R_\pm}
\def\Msun{M_\odot}

\def\pphi{{p_{\varphi}^0}}
\def\CE{{\rm CE_0}}
\def\UE{{\rm UE_0}}

\def\TEOBResumS{\texttt{TEOBResumS}}
\def\TEOBResumSHyp{\texttt{TEOBResumSHyp}}
\def\TEOBResumSP{\texttt{TEOBResumSP}}
\def\TEOBResumShm{\texttt{TEOBiResumS\_SM}}
\def\TEOBResumSecce{\texttt{TEOBResumSecce}}
\def\TEOBResumROM{\texttt{TEOBResum\_ROM}}
\def\SEOBNRvq{{\texttt{SEOBNRv4}}}
\def\SEOBNRvqT{{\texttt{SEOBNRv4T}}}
\def\CC{{C\nolinebreak[4]\hspace{-.05em}\raisebox{.4ex}{\tiny\bf ++}}}

\newcommand{\an}[1]{{\textcolor{red}{\texttt{AN: #1}} }}
\newcommand{\pr}[1]{{\textcolor{cyan}{\texttt{PR: #1}} }}
\newcommand{\bs}[1]{{\textcolor{blue}{\texttt{SB: #1}} }}
\newcommand{\mb}[1]{{\textcolor{purple}{\texttt{MB: #1}} }}
\newcommand{\rg}[1]{{\textcolor{green}{\texttt{RG: #1}} }}
\newcommand{\gc}[1]{{\textcolor{violet}{\texttt{GC: #1}} }}
\usepackage[normalem]{ulem} 
\newcommand{\oldtxt}[1]{{\sout{#1}}}
\newcommand{\oldnewtxt}[2]{{\sout{#1}{\red{{#2}}}}}
\newcommand{\newtxt}[1]{{\textcolor{magenta}{{\bf[new]} #1}}}
\newcommand{\red}[1]{{\textcolor{red}{#1}}}

\setcounter{equation}{0}
\setcounter{figure}{0}
\setcounter{table}{0}
\setcounter{page}{1}
\makeatletter
\renewcommand{\theequation}{S\arabic{equation}}
\renewcommand{\figurename}{SUPPLEMENTARY FIGURE}
\renewcommand{\tablename}{SUPPLEMENTARY TABLE}

\begin{document}

\title{Supplemental material}
\author{Rossella \surname{Gamba}${}^{1}$}
\author{Matteo \surname{Breschi}${}^{1}$}
\author{Gregorio \surname{Carullo}${}^{2,3}$}
\author{Piero \surname{Rettegno}${}^{4,5}$}
\author{Simone \surname{Albanesi}${}^{4,5}$}
\author{Sebastiano \surname{Bernuzzi}${}^{1}$}
\author{Alessandro \surname{Nagar}${}^{5,6}$\footnote{Corresponding author: alessandro.nagar@gmail.com}}
\affiliation{${}^{1}$ Theoretisch-Physikalisches Institut, Friedrich-Schiller-Universit{\"a}t Jena, 07743, Jena, Germany}
\affiliation{${}^{2}$ Dipartimento di Fisica ``Enrico Fermi'', Universit\`a di Pisa, Pisa I-56127, Italy}
\affiliation{${}^{3}$ INFN sezione di Pisa, Pisa I-56127, Italy}
\affiliation{${}^{4}$ Dipartimento di Fisica, Universit\`a di Torino, via P. Giuria 1, 10125 Torino, Italy}
\affiliation{${}^{5}$ INFN Sezione di Torino, Via P. Giuria 1, 10125 Torino, Italy}
\affiliation{${}^{6}$ Institut des Hautes Etudes Scientifiques, 91440 Bures-sur-Yvette, France}

\maketitle

\fi

\subsection{Validation of the dynamical capture model}

\begin{figure}
\label{fig:mismatch_ecc}
\centering
\includegraphics[width=0.5\textwidth]{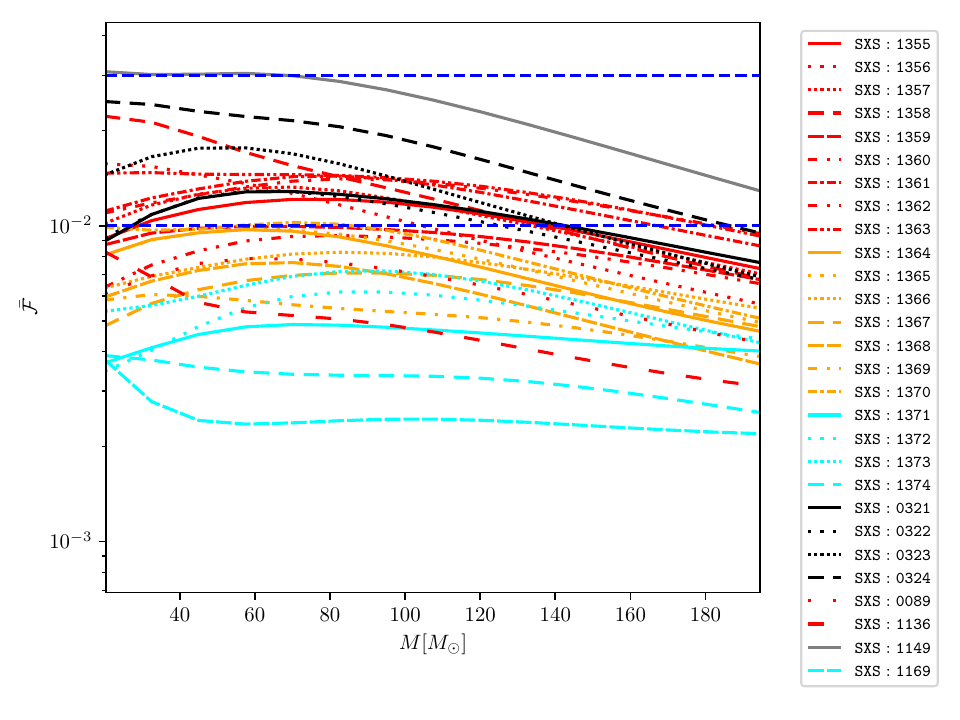}
\caption{{\bf EOB/NR unfaithfulness for mildly eccentric systems}. Comparison between 28 mildly eccentric NR simulations from 
the SXS catalog\cite{SXS:catalog} and the eccentric $\TEOBResumS$. These mismatches improve and extend those 
computed in previous work\cite{Chiaramello:2020ehz}.}
\end{figure}

The EOB\cite{Buonanno:1998gg}${}^{,}$\citesupp{Buonanno:2000ef,Damour:2000we,Damour:2001tu,Damour:2015isa} waveform model $\TEOBResumS$ employed in our study 
is based on an highly accurate approximant for quasi-circular binaries \citesupp{Nagar:2018zoe, Nagar:2019wds, Nagar:2020pcj, Riemenschneider:2021ppj}. The eccentric model
has been extensively tested in previous works\cite{Chiaramello:2020ehz, Albanesi:2021rby}
via mismatch and waveform phasing comparisons. In detail, the model has been compared to:
\begin{enumerate}
 \item[(i)] 28 equal-mass mildly eccentric NR simulations\cite{Chiaramello:2020ehz}  from the Simulating 
 eXtreme Spacetimes (SXS) collaboration\citesupp{Chu:2009md,Lovelace:2010ne,Lovelace:2011nu,Buchman:2012dw, Hemberger:2013hsa,
 Scheel:2014ina,Blackman:2015pia, Lovelace:2014twa,Mroue:2013xna,Kumar:2015tha,Chu:2015kft, Boyle:2019kee,SXS:catalog}.
 The comparison involved either time-domain phasing analysis or EOB/NR unfaithfulness computations. 
 For completeness, this second analysis is repeated here with the implementation of the model employed in the analysis,
 and illustrated in Fig.~\ref{fig:mismatch_ecc}. The model shows robust mismatches well below the $3\%$ threshold 
 for system with masses ranging from $20 \Msun$ to $200 \Msun$ with the Advanced LIGO design sensitivity curve\citesupp{aLIGODesign_PSD}.
 
 \item[(ii)] 10 equal-mass scattering simulations from\cite{Damour:2014afa} used to check the scattering angle, but for which 
                waveforms are not available\cite{Nagar:2020xsk}. 
   
 \item[(iii)]  111 waveforms generated by a nonspinning 
 test particle along planar geodesics in Kerr spacetime with eccentricities up to 
 0.9 and dimensionless Kerr spin magnitude up to $\hat{a}=0.9$. The analysis was also extended 
 to nongeodesic motion considering the transition from inspiral to plunge driven by the 
 EOB radiation reaction force considered in this work. Dynamical capture scenarios on
 a Schwarzschild spacetime were also considered. The accuracy of the fluxes has been studied 
 in Sec.~IV of Albanesi et al.\cite{Albanesi:2021rby}, while the waveform has been analyzed 
 in Sec.~V. In particular, the analytical/numerical comparisons for the non-geodesics configurations 
 can be found in Fig.~13 and  Fig.~14 therein.
  
\end{enumerate}

These tests demonstrate the goodness of our model in a regime which is, in principle, different from the one explored here, where an almost equal-mass system undergoes a dynamical capture. Notably, however, some important insight can be extrapolated from this information. 
Firstly, we observe that although the use of a circular ringdown is an approximation, the tests against test-mass waveforms from dynamical 
encounters show a good performance for the configurations that circularize during the last encounter, i.e. those selected by the parameter estimation,
see in particular Fig.~14 of\cite{Albanesi:2021rby}.
The region where the circular ringdown performs less well is the direct-capture and head-on scenario (also expected on physical grounds). 
However, these configurations are also excluded from the parameter estimation by their smaller likelihood values.
Secondly, the comparison with eccentric comparable mass data proves that the radiation reaction employed is highly accurate in that regime. 
Finally, the EOB/NR comparisons of the scattering angle of relativistic equal-mass BBH is a strong test of the dynamics (of both the conservative and dissipative sector), 
that probes the model in a very challenging physical scenario\cite{Nagar:2020xsk}.

\begin{figure}[t]
\begin{center}
\includegraphics[width=0.5\textwidth]{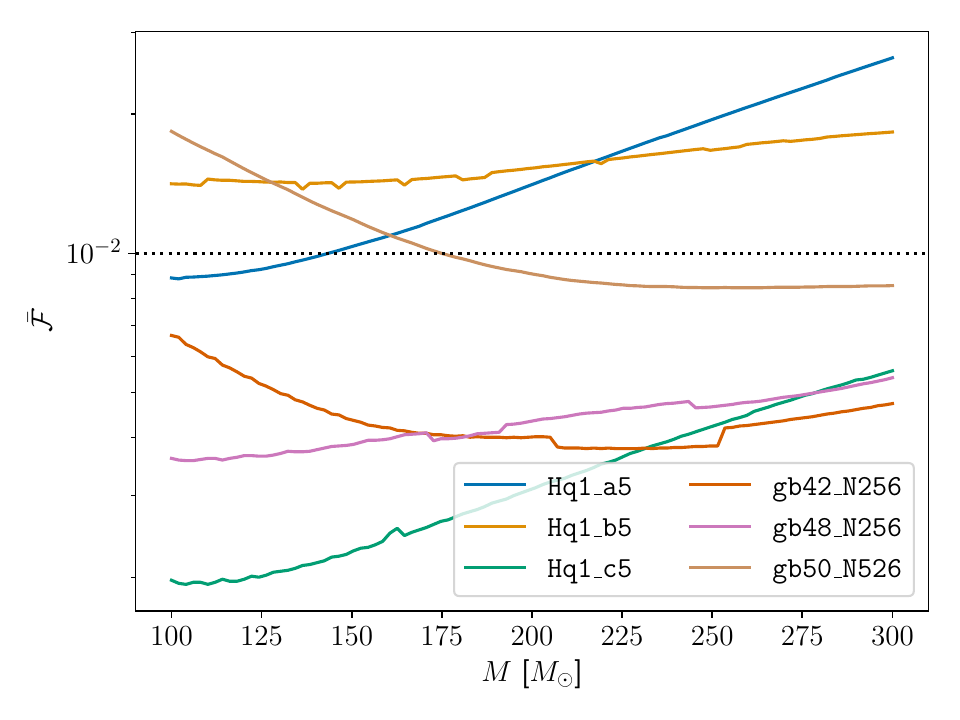}
\caption{{\bf EOB/NR unfaithfulness for highly eccentric and capture configurations}. 
Comparison between the NR simulations of Table~\ref{tab:nrhyp}, produced for this work, 
and the eccentric $\TEOBResumS$. We consider frequencies between $11$ and $512$ Hz, 
use the Hanford PSD of GW190521 and compute $\bar{\mathcal{F}}$ for systems 
with masses $M \in [100, 300] M_{\odot}$.}
\label{fig:mismatch_hyp}
\end{center}
\end{figure}

\begin{figure}[t]
\begin{center}
\includegraphics[width=0.3\textwidth]{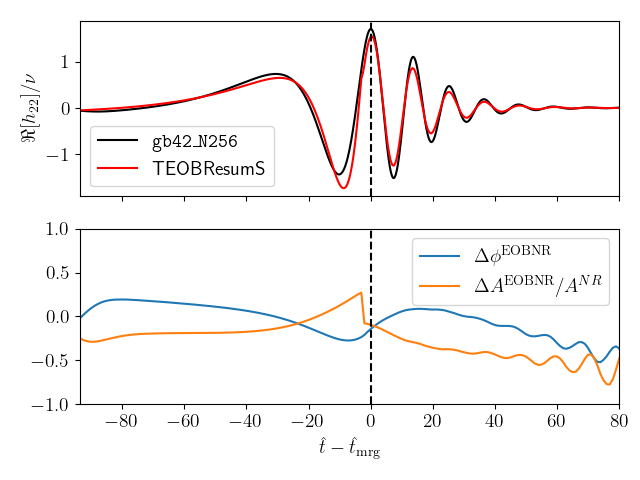}
\includegraphics[width=0.3\textwidth]{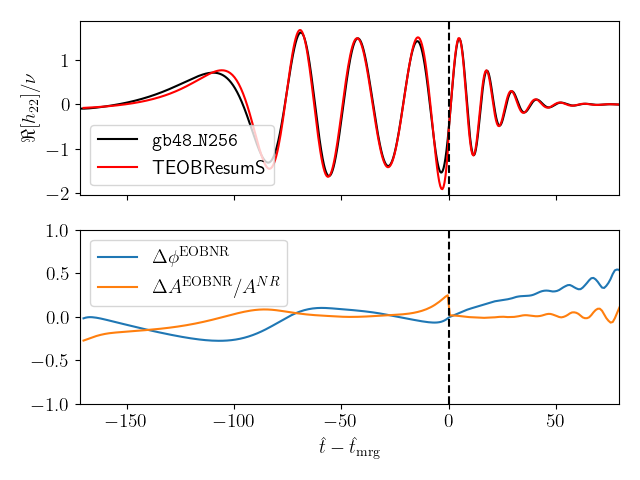}
\includegraphics[width=0.3\textwidth]{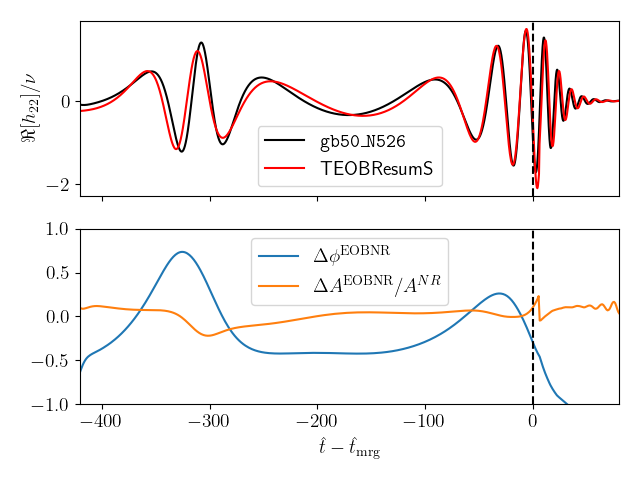}
\includegraphics[width=0.3\textwidth]{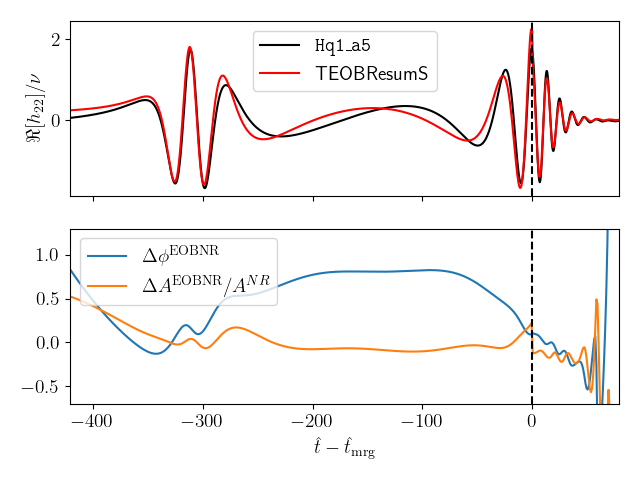} 
\includegraphics[width=0.3\textwidth]{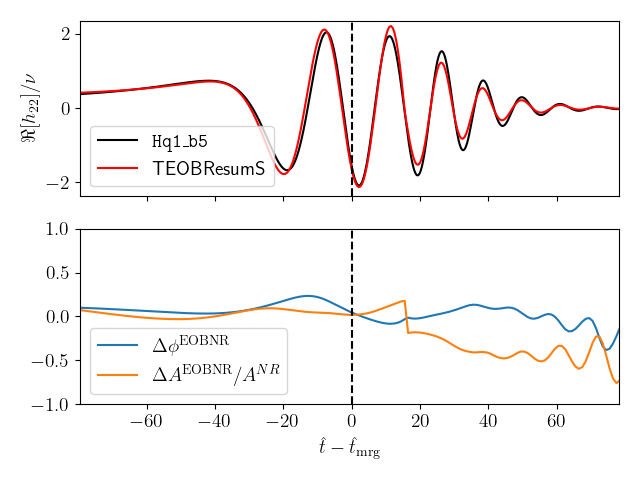}
\includegraphics[width=0.3\textwidth]{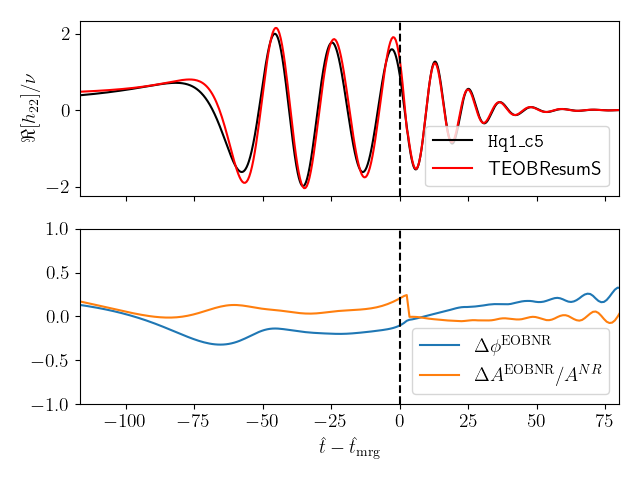}
\caption{{\bf Time-domain EOB/NR phasing comparison}. Comparing the Numerical Relativity waveforms of Table~\ref{tab:nrhyp} 
to the analytical ones obtained using the eccentric  \TEOBResumS{}. 
For each configuration, the top panel displays the amplitude and the real part of the dominant multipole $h_{22}$, while the bottom panel 
shows the phase difference $\Delta\phi^{\rm EOBNR} = \phi^{\rm EOB} - \phi^{\rm NR}$ and the relative amplitude difference 
$\Delta A^{\rm EOBNR}/A^{\rm NR} = (A^{\rm EOB} - A^{\rm NR})/A^{\rm NR}$. Despite the lack of NQC corrections or of an 
hyperbolic-NR-informed ringdown, \TEOBResumS{} quantitatively captures the NR waveform for any configuration.}
\label{fig:phasing_hyp}
\end{center}
\end{figure}

To further corroborate these observations in the regime of direct interest for this publication, we produced six equal-mass, nonspinning 
simulations of highly eccentric systems or dynamical captures using the NR code ${\tt GR-Athena++}$~\citesupp{Daszuta:2021ecf}, see Table~\ref{tab:nrhyp}.
Three simulations reproduce configurations of Gold and Bruegmann~\cite{Gold:2012tk} ($\tt{gb42\_N256}$, $\tt{gb48\_N256}$ and $\tt{gb50\_N256}$); 
the three remaining ones are instead completely new configurations ($\tt{Hq1\_a5}$, $\tt{Hq1\_b5}$ and $\tt{Hq1\_c5}$) with initial conditions chosen 
to target the parameter space selected by our GW190521 analysis.
In order to compare NR and EOB waveforms, one needs consistent initial energy, angular momentum and separation.
While the first two quantities are in principle gauge-invariant, to get the latter we need to
convert from Arnowitt-Deser-Misner (ADM) coordinates to EOB coordinates using a 
2PN-accurate transformation~\cite{Buonanno:1998gg}\citesupp{Bini:2012ji}.
However, the existence of NR junk radiation, resolution effects as well as the finite 
PN order of the ADM to EOB transformation can cause small differences between the 
NR and EOB initial data. While small variations in the energy and angular can significantly 
change the phenomenology
of the waveform, as shown in Fig.~2 of\cite{Nagar:2020xsk}, small inaccuracies 
in the initial separation are not relevant as long as the bodies are initially far enough. 
In this scenario, the effect of the radiation reaction is negligible at the beginning of the evolution,
and small shifts in initial separation correspond to global constant time shifts.
As such, in order to estimate the ``optimal'' values of $(\hat{E}_0, \hat{p}_\varphi^0)$, we first additionally minimized 
the mismatch on the initial energy and angular momentum over a small interval around the values extracted from the procedure described above,
allowing a relative error up to $1\%$ in energy and up to $6\%$ in angular momentum.
This procedure was performed only for a single reference value of total mass 
($M = 250 M_{\odot}$, i.e. the detector frame mass of GW190521), using the expression:
\begin{equation}
\label{eq:mismatch}
\bar{\mathcal{F}} = 1 - \max_{t_0, \phi_0, \hat{E}_0, \hat{p}_{\varphi}^0} \frac{(h^{\rm NR}, h^{\rm EOB})}{\sqrt{(h^{\rm NR}, h^{\rm NR})(h^{\rm EOB}, h^{\rm EOB})}} \, ,
\end{equation}
where $(\cdot, \cdot)$ denotes the usual noise weighted inner product
\begin{equation*}
(a, b) = 4 \Re \int_{f_0}^{f_{max}}\frac{\tilde{a}(f) \tilde{b}(f)}{S_n(f)} df \, ,
\end{equation*}
and $S_n$ is the power spectral density (PSD) of the detector, in this case chosen to be GW190521 Hanford's PSD.
The initial conditions found with this procedure are then employed to compute $\bar{\mathcal{F}}$ for all values of the total mass $M$ 
considered using the standard definition of mismatch (the same as above, without varying the initial energy and angular momentum).
We considered frequencies between $11$ and $512$ Hz and total masses $M \in [100, 300] M_{\odot}$. 
We found mismatches between $0.2\%$ and $3\%$.
Details about the simulations are listed in Table~\ref{tab:nrhyp}, while the results of the computation can be seen in
Fig.~\ref{fig:mismatch_hyp}. The corresponding time-domain EOB/NR phase comparisons of the $\ell=m=2$ waveform 
are shown in Fig.~\ref{fig:phasing_hyp}. 
Let us recall that we use the following multipolar decomposition of the strain waveform
\be
h_+ - i h_\times = D_{L}^{-1} \sum_{\ell=2}^{\rm \ell_{\rm max}}\sum_{m=-\ell}^{\ell}h_{\lm}\,{}_{-2}Y_\lm \ ,
\ee
where $D_L$ is the luminosity distance and  ${}_{-2}Y_\lm$ are the $s=-2$ spin-weighted spherical harmonics.
Focusing only on the $\ell=m=2$ dominant mode, the waveform is decomposed in amplitude and phase as $h_{22}(t)=A(t)e^{-i\phi(t)}$. 
For each configuration in Fig.~\ref{fig:phasing_hyp} we compare the real part of the EOB and NR waveform 
and explicitly show the phase difference $\Delta\phi^{\rm EOBNR}\equiv \phi^{\rm EOB}-\phi^{\rm NR}$ and the relative amplitude
difference $\Delta A^{\rm EOBNR}/A^{\rm NR}\equiv |A^{\rm EOB}-A^{\rm NR}|/A^{\rm NR}$.
Note that the NR-informed quasi-circular ringdown\cite{Nagar:2020pcj} delivers rather faithful representation
of the NR phasing, while the amplitude might be underestimated. This is consistent with the findings in
the test-particle limit\cite{Albanesi:2021rby}. 

\begin{figure}[t]
\begin{center}
\includegraphics[width=0.8\textwidth]{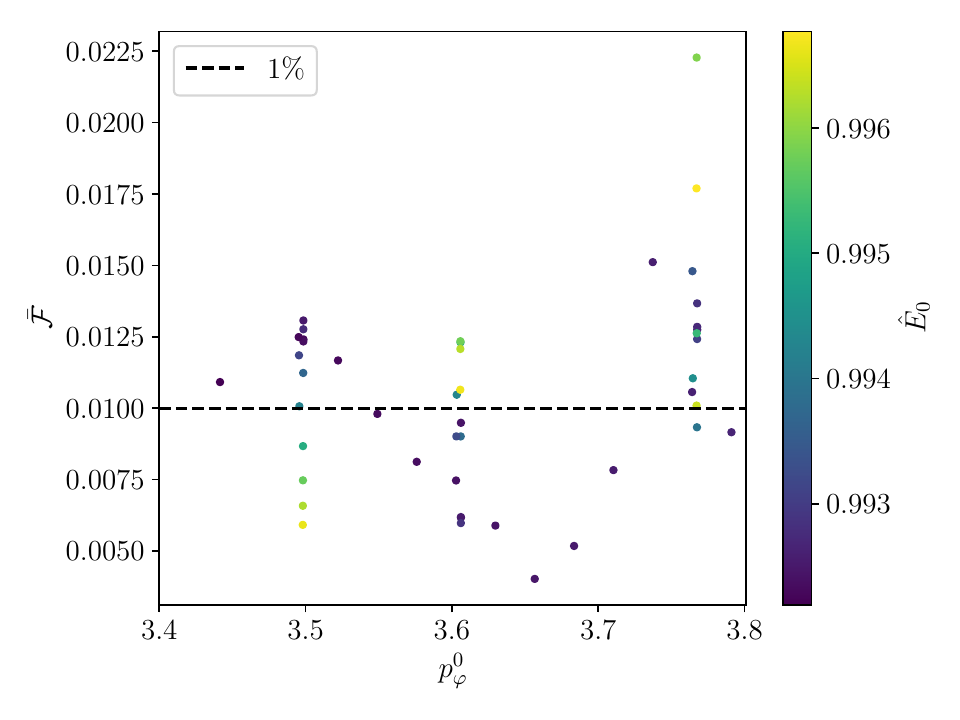}
\caption{{\bf EOB/NR unfaithfulness for the highly eccentric NR waveforms from the RIT catalog.}
We compare 46 nonspinning highly eccentric datasets from RIT\cite{Healy:2022wdn} with the eccentric \TEOBResumS{} model.
 We select only those systems whose best matching EOB initial conditions, estimated via Eq.~\eqref{eq:mismatch}, fall inside the posterior samples. 
 We consider frequencies between $11$ and $512$ Hz, use the Hanford PSD of GW190521 and 
 compute $\bar{\mathcal{F}}$ with fixed total mass $M = 250 M_{\odot}$.}
\label{fig:mismatch_rit}
\end{center}
\end{figure}

During the development of this work, the RIT group released the data of the large number of NR simulations 
of highly eccentric BBH systems\cite{Healy:2022wdn} used in the analysis of GW190521 of\cite{Gayathri:2020coq}. 
Figure~\ref{fig:mismatch_rit} shows the unfaithfulness of our model computed against all such simulations with initial eccentricity 
larger than $0.5$, zero spins, $q\leq 8$ and initial angular momentum and energy consistent with our priors ($p_\varphi \geq 3.4$ at initial separation $r \sim 20$).
Notably, the 46 simulations selected display typical EOB/NR mismatches below $3\%$, with  about half of them more than $99\%$ faithful to NR.

Additional detailed EOB/NR comparisons considering a larger number of simulations, an improved EOB model\citesupp{Nagar:2021xnh,Hopper:2022rwo}, a wider parameter space as well as higher modes 
and spin effects will be presented in a future work (Andrade et. al).

\begin{table}
\begin{center}
\begin{tabular}{c|c|c|c|c|c|c|c} 
 \hline 
 \hline
ID  & $\hat{r}_0^{\rm EOB}$ & $\hat{E}^{\rm EOB}$ & $\hat{p}_\varphi^{\rm EOB}$ & $\hat{E}^{\rm EOB}_{\rm opt}$ & $\hat{p}_{\varphi\; \rm opt}^{\rm EOB}$&$\Delta E/E[\%]$ & $\Delta p_{\varphi}/p_{\varphi}[\%]$\\ 
 \hline 
$\tt Hq1\_a5$ & $100.000$ & $1.003$ & $3.970$ & $1.005$ & $3.987$ & $-0.132$ & $-0.440$ \\ 
$\tt Hq1\_b5$ & $100.000$ & $1.008$ & $3.970$ & $1.010$ & $4.049$ & $-0.192$ & $-2.000$ \\ 
$\tt Hq1\_c5$ & $100.000$ & $1.007$ & $3.970$ & $1.015$ & $4.181$ & $-0.799$ & $-5.323$ \\ 
$\tt gb42\_N256$ & $20.842$ & $0.994$ & $3.305$ & $0.994$ & $3.238$ & $0.018$ & $2.033$ \\ 
$\tt gb48\_N256$ & $20.869$ & $0.994$ & $3.671$ & $0.997$ & $3.762$ & $-0.267$ & $-2.476$ \\ 
$\tt gb50\_N526$ & $20.878$ & $0.994$ & $3.784$ & $0.993$ & $3.771$ & $0.131$ & $0.342$\\ 
\hline
\hline
\end{tabular}
\end{center}
\label{tab:nrhyp}
\caption{{\bf EOB initial data to match NR simulations}. The values correspond to the equal-mass, nonspinning NR 
data of Figs.~\ref{fig:mismatch_hyp} and~\ref{fig:phasing_hyp}. The first column of the table reports the configuration 
label. The quantities $\hat{r}_0^{\rm EOB}$,$\hat{E}^{\rm EOB}$ and 
$\hat{p}_\varphi^{\rm EOB}$ are the initial EOB radial separation, initial energy and initial angular momentum 
obtained by mapping the initial position and momenta of the punctures into EOB 
coordinates via the 2PN coordinate transformation between ADM and EOB coordinates\cite{Bini:2012ji}. 
The quantities $\hat{E}^{\rm EOB}_{\rm opt}$ and $\hat{p}_{\varphi\;\rm opt}^{\rm EOB}{}$, instead, are
the corresponding values which minimize the EOB/NR mismatch at $M=250 M_{\odot}$. 
The last two columns report the corresponding relative differences.}
\end{table}

\subsection{Injection-recovery studies}

To better understand the increase in $\log(\mathcal{L})_{\rm max}$ and Bayes' factor observed in the 
main text we perform two additional "injection-recovery" studies:
\begin{enumerate}
\item[(i)] Using the {\tt NRSur7dq4} waveform model, we simulate a quasicircular signal with the maximum 
likelihood parameters recovered from GW190521 into gaussian noise generated with the PSD estimated close to the event, and recover it with both {\tt NRSur7dq4} and ${\tt TEOBResumS}$ 
\item[(ii)] We perform a self-consistency test: we simulate and recover a signal with the hyperbolic model employed in the main text, ${\tt TEOBResumS}$, with parameters similar to GW190521.
\end{enumerate}
In all cases we adopt the same PE settings and priors as the ones used in the analysis shown in the main paper. 

Our results for the first injection-recovery test are reported in Table~\ref{tab:injections_nrsur}. 
Both the maximum likelihood and the Bayes Factor recovered with {\tt NRSur7dq4} are slightly larger than those obtained with ${\tt TEOBResumS}$. 
On the one hand, this indicates that the significant increase in Bayes Factor and likelihood we observe in our analyses is not obtained when the real signal is generated by a precessing, quasi circular source. 
On the other hand, it also shows how short-lived precessing signals
can be matched reasonably well also by dynamical-capture waveform models, 
i.e. the symmetric scenario with respect to the one discussed in\cite{CalderonBustillo:2020srq}.
Note that, due to the specific noise realization, at times the injected parameters lie just outside the $90\%$ credible levels even when the recovery is performed with the surrogate.

Table~\ref{tab:injections_teob}, instead, displays the injected and recovered parameters obtained from our second injection-recovery test.
All the recovered parameters lie, as expected, well within the $90\%$ intervals. This indicates that our inference framework (including the model implementation) behaves correctly, and is able to recover the simulated parameters in spite of the complicated structure of the $(E,p_\varphi)$ parameter space.

\begin{table}
\begin{center}
\resizebox{0.7\textwidth}{!}{\begin{tabular}{lccc}
\hline
\hline
& Injected ({\tt NRSur7dq4}) & {\tt NRSur7dq4} &  {\tt TEOBResumS}\\
\hline
$m_1$ [$\Msun$]&  $135$ & $155^{+ 19}_{-19}$ & $145^{+ 21}_{-20}$\\
$m_2$ [$\Msun$]&  $125$ & $ 89^{+ 36}_{-34}$ & $ 88^{+ 28}_{-36}$ \\
$M$ [$\Msun$]  & $260$ & $244^{+ 38}_{-36}$ & $232^{+ 15}_{-20}$ \\
$m_1^{\rm source}$ [$\Msun$]  & $90.01 $ & $92 ^{+ 19}_{- 20}$ & $72^{+ 26}_{-14}$\\
$m_2^{\rm source}$ [$\Msun$]  & $ 83.397 $ & $52^{+ 21}_{- 18}$ & $43^{+ 17}_{-16}$\\
$M_{\rm source}$ [$\Msun$]  & $173.40$ & $141^{+ 31}_{-21}$ & $115^{+ 29}_{-16}$ \\
$m_2/m_1\leq 1$  & $0.92$ & $0.57^{+ 0.28}_{-0.23}$ & $0.61^{+ 0.31}_{- 0.29 }$ \\
$\chi_{\rm eff}$ & $-0.29$ & $-0.18 ^{+ 0.43 }_{- 0.41 }$ & -- \\
$\chi_{\rm p}$   & $0.79$ &  $0.51^{+ 0.33}_{-0.38}$ & --\\
$E^0/M$ & --  & -- & $1.010^{+ 0.013}_{- 0.009}$ \\
$\pphi$ & -- & -- & $4.08^{+ 0.32}_{- 0.27}$ \\
$D_L$ [Mpc] & $2935$ & $4390^{+ 3370}_{- 2060}$& $6950^{+ 2590}_{-3290} $ \\
$\log({\cal L})_{\rm max}$  & -- &  $56.70$ &  $52.84$ \\
$\log{\cal B}^{\rm signal}_{\rm noise}$ & -- & $29.65$ & $28.43$\\  
\hline
\hline
\end{tabular}}
\end{center}
\caption{\label{tab:injections_nrsur}
{\bf First  injection-recovery study.} Results of our ${\tt NRSur7dq4}-{\tt NRSur7dq4}$ and ${\tt NRSur7dq4}-{\tt TEOBResumS}$ injection-recovery studies 
in GW190521 gaussian noise. Posteriors are expressed, as is standard, via their median and $90\%$ credible intervals. Notably, the $\log({\cal L})_{\rm max}$ and $\log{\cal B}^{\rm signal}_{\rm noise}$ recovered with the NR surrogate are higher than those obtained with the hyperbolic model.}
\end{table}

\begin{table}
\begin{center}
\resizebox{0.7\textwidth}{!}{\begin{tabular}{lcc}
\hline
\hline
& Injected ({\tt TEOBResumS}) & {\tt TEOBResumS}\\
\hline
$m_1^{\rm source}$ [$\Msun$]  & $81$  & $93^{+ 22 }_{- 14 }$ \\
$m_2^{\rm source}$ [$\Msun$]  & $71$  & $60 ^{+ 14 }_{- 18 }$  \\
$M_{\rm source}$ [$\Msun$]    & $152$ & $153^{+ 16 }_{- 10 }$ \\
$m_2/m_1\leq 1$  & $0.88$ & $0.64^{+ 0.25}_{- 0.27 }$ \\
$E^0/M$ & $1.012$  & $1.010 ^{+ 0.007 }_{- 0.008 }$\\
$\pphi$ & $4.21$ & $4.04 ^{+ 0.27}_{- 0.29}$ \\
$D_L$ [Mpc] & $3000$ &  $2400 ^{+ 790}_{- 930}$\\
$\log({\cal L})_{\rm max}$  & -- & $316.84$  \\
$\log{\cal B}^{\rm signal}_{\rm noise}$ & -- & $282.44$ \\  
\hline
\hline
\end{tabular}}
\end{center}
\caption{\label{tab:injections_teob}
{\bf Second  injection-recovery study.} Results of our ${\tt TEOBResumS}-{\tt TEOBResumS}$ injection-recovery 
experiment in GW190521 gaussian noise. Posteriors are expressed, as is standard, via their median and $90\%$ credible intervals. 
The injected true values all fall within the $90\%$ intervals.}
\end{table}

\bibliographystylesupp{plain}
\bibliographysupp{supp.bbl}

\fi


\begin{thebibliography}{10}

\bibitem{SXS:catalog}
{SXS Gravitational Waveform Database}.
\newblock \url{https://data.black-holes.org/waveforms/index.html}.

\bibitem{aLIGODesign_PSD}
{Updated Advanced LIGO sensitivity design curve}.
\newblock \url{https://dcc.ligo.org/LIGO-T1800044/public}.

\bibitem{Albanesi:2021rby}
Simone Albanesi, Alessandro Nagar, and Sebastiano Bernuzzi.
\newblock {Effective one-body model for extreme-mass-ratio spinning binaries on
  eccentric equatorial orbits: Testing radiation reaction and waveform}.
\newblock {\em Phys. Rev. D}, 104(2):024067, 2021.

\bibitem{Bini:2012ji}
Donato Bini and Thibault Damour.
\newblock {Gravitational radiation reaction along general orbits in the
  effective one-body formalism}.
\newblock {\em Phys.Rev.}, D86:124012, 2012.

\bibitem{Blackman:2015pia}
Jonathan Blackman, Scott~E. Field, Chad~R. Galley, Béla Szilágyi, Mark~A.
  Scheel, Manuel Tiglio, and Daniel~A. Hemberger.
\newblock {Fast and Accurate Prediction of Numerical Relativity Waveforms from
  Binary Black Hole Coalescences Using Surrogate Models}.
\newblock {\em Phys. Rev. Lett.}, 115(12):121102, 2015.

\bibitem{Boyle:2019kee}
Michael Boyle et~al.
\newblock {The SXS Collaboration catalog of binary black hole simulations}.
\newblock {\em Class. Quant. Grav.}, 36(19):195006, 2019.

\bibitem{Buchman:2012dw}
Luisa~T. Buchman, Harald~P. Pfeiffer, Mark~A. Scheel, and Bela Szilagyi.
\newblock {Simulations of non-equal mass black hole binaries with spectral
  methods}.
\newblock {\em Phys. Rev.}, D86:084033, 2012.

\bibitem{Buonanno:1998gg}
A.~Buonanno and T.~Damour.
\newblock {Effective one-body approach to general relativistic two-body
  dynamics}.
\newblock {\em Phys. Rev.}, D59:084006, 1999.

\bibitem{Buonanno:2000ef}
Alessandra Buonanno and Thibault Damour.
\newblock {Transition from inspiral to plunge in binary black hole
  coalescences}.
\newblock {\em Phys. Rev.}, D62:064015, 2000.

\bibitem{CalderonBustillo:2020srq}
Juan~Calder\'on Bustillo, Nicolas Sanchis-Gual, Alejandro Torres-Forn\'e,
  Jos\'e~A. Font, Avi Vajpeyi, Rory Smith, Carlos Herdeiro, Eugen Radu, and
  Samson H.~W. Leong.
\newblock {GW190521 as a Merger of Proca Stars: A Potential New Vector Boson of
  $8.7\times 10^{-13}$ eV}.
\newblock {\em Phys. Rev. Lett.}, 126(8):081101, 2021.

\bibitem{Chiaramello:2020ehz}
Danilo Chiaramello and Alessandro Nagar.
\newblock {Faithful analytical effective-one-body waveform model for
  spin-aligned, moderately eccentric, coalescing black hole binaries}.
\newblock {\em Phys. Rev. D}, 101(10):101501, 2020.

\bibitem{Chu:2015kft}
Tony Chu, Heather Fong, Prayush Kumar, Harald~P. Pfeiffer, Michael Boyle,
  Daniel~A. Hemberger, Lawrence~E. Kidder, Mark~A. Scheel, and Bela Szilagyi.
\newblock {On the accuracy and precision of numerical waveforms: Effect of
  waveform extraction methodology}.
\newblock {\em Class. Quant. Grav.}, 33(16):165001, 2016.

\bibitem{Chu:2009md}
Tony Chu, Harald~P. Pfeiffer, and Mark~A. Scheel.
\newblock {High accuracy simulations of black hole binaries:spins anti-aligned
  with the orbital angular momentum}.
\newblock {\em Phys. Rev.}, D80:124051, 2009.

\bibitem{Damour:2001tu}
Thibault Damour.
\newblock {Coalescence of two spinning black holes: An effective one- body
  approach}.
\newblock {\em Phys. Rev.}, D64:124013, 2001.

\bibitem{Damour:2014afa}
Thibault Damour, Federico Guercilena, Ian Hinder, Seth Hopper, Alessandro
  Nagar, et~al.
\newblock {Strong-Field Scattering of Two Black Holes: Numerics Versus
  Analytics}.
\newblock 2014.

\bibitem{Damour:2000we}
Thibault Damour, Piotr Jaranowski, and Gerhard Schaefer.
\newblock {On the determination of the last stable orbit for circular general
  relativistic binaries at the third postNewtonian approximation}.
\newblock {\em Phys. Rev.}, D62:084011, 2000.

\bibitem{Damour:2015isa}
Thibault Damour, Piotr Jaranowski, and Gerhard Schäfer.
\newblock {Fourth post-Newtonian effective one-body dynamics}.
\newblock {\em Phys. Rev.}, D91(8):084024, 2015.

\bibitem{Daszuta:2021ecf}
Boris Daszuta, Francesco Zappa, William Cook, David Radice, Sebastiano
  Bernuzzi, and Viktoriya Morozova.
\newblock {GR-Athena++: Puncture Evolutions on Vertex-centered Oct-tree
  Adaptive Mesh Refinement}.
\newblock {\em Astrophys. J. Supp.}, 257(2):25, 2021.

\bibitem{Gayathri:2020coq}
V.~Gayathri, J.~Healy, J.~Lange, B.~O'Brien, M.~Szczepanczyk, Imre Bartos,
  M.~Campanelli, S.~Klimenko, C.~O. Lousto, and R.~O'Shaughnessy.
\newblock {Eccentricity estimate for black hole mergers with numerical
  relativity simulations}.
\newblock {\em Nature Astron.}, 6(3):344--349, 2022.

\bibitem{Gold:2012tk}
Roman Gold and Bernd Br{\"u}gmann.
\newblock {Eccentric black hole mergers and zoom-whirl behavior from elliptic
  inspirals to hyperbolic encounters}.
\newblock {\em Phys. Rev.}, D88(6):064051, 2013.

\bibitem{Healy:2022wdn}
James Healy and Carlos~O. Lousto.
\newblock {The Fourth RIT binary black hole simulations catalog: Extension to
  Eccentric Orbits}.
\newblock 1 2022.

\bibitem{Hemberger:2013hsa}
Daniel~A. Hemberger, Geoffrey Lovelace, Thomas~J. Loredo, Lawrence~E. Kidder,
  Mark~A. Scheel, Béla Szilágyi, Nicholas~W. Taylor, and Saul~A. Teukolsky.
\newblock {Final spin and radiated energy in numerical simulations of binary
  black holes with equal masses and equal, aligned or anti-aligned spins}.
\newblock {\em Phys. Rev.}, D88:064014, 2013.

\bibitem{Hopper:2022rwo}
Seth Hopper, Alessandro Nagar, and Piero Rettegno.
\newblock {Strong-field scattering of two spinning black holes: Numerics versus
  Analytics}.
\newblock 4 2022.

\bibitem{Kumar:2015tha}
Prayush Kumar, Kevin Barkett, Swetha Bhagwat, Nousha Afshari, Duncan~A. Brown,
  Geoffrey Lovelace, Mark~A. Scheel, and Béla Szilágyi.
\newblock {Accuracy and precision of gravitational-wave models of inspiraling
  neutron star-black hole binaries with spin: Comparison with matter-free
  numerical relativity in the low-frequency regime}.
\newblock {\em Phys. Rev.}, D92(10):102001, 2015.

\bibitem{Lovelace:2011nu}
Geoffrey Lovelace, Michael Boyle, Mark~A. Scheel, and Bela Szilagyi.
\newblock {Accurate gravitational waveforms for binary-black-hole mergers with
  nearly extremal spins}.
\newblock {\em Class. Quant. Grav.}, 29:045003, 2012.

\bibitem{Lovelace:2014twa}
Geoffrey Lovelace et~al.
\newblock {Nearly extremal apparent horizons in simulations of merging black
  holes}.
\newblock {\em Class. Quant. Grav.}, 32(6):065007, 2015.

\bibitem{Lovelace:2010ne}
Geoffrey Lovelace, Mark.A. Scheel, and Bela Szilagyi.
\newblock {Simulating merging binary black holes with nearly extremal spins}.
\newblock {\em Phys.Rev.}, D83:024010, 2011.

\bibitem{Mroue:2013xna}
Abdul~H. Mroue, Mark~A. Scheel, Bela Szilagyi, Harald~P. Pfeiffer, Michael
  Boyle, et~al.
\newblock {A catalog of 174 binary black-hole simulations for
  gravitational-wave astronomy}.
\newblock {\em Phys.Rev.Lett.}, 111:241104, 2013.

\bibitem{Nagar:2018zoe}
Alessandro Nagar et~al.
\newblock {Time-domain effective-one-body gravitational waveforms for
  coalescing compact binaries with nonprecessing spins, tides and self-spin
  effects}.
\newblock {\em Phys. Rev.}, D98(10):104052, 2018.

\bibitem{Nagar:2019wds}
Alessandro Nagar, Geraint Pratten, Gunnar Riemenschneider, and Rossella Gamba.
\newblock {A Multipolar Effective One Body Model for Non-Spinning Black Hole
  Binaries}.
\newblock 2019.

\bibitem{Nagar:2021xnh}
Alessandro Nagar and Piero Rettegno.
\newblock {The next generation: Impact of high-order analytical information on
  effective one body waveform models for noncircularized, spin-aligned black
  hole binaries}.
\newblock 8 2021.

\bibitem{Nagar:2020xsk}
Alessandro Nagar, Piero Rettegno, Rossella Gamba, and Sebastiano Bernuzzi.
\newblock {Effective-one-body waveforms from dynamical captures in black hole
  binaries}.
\newblock {\em Phys. Rev. D}, 103(6):064013, 2021.

\bibitem{Nagar:2020pcj}
Alessandro Nagar, Gunnar Riemenschneider, Geraint Pratten, Piero Rettegno, and
  Francesco Messina.
\newblock {Multipolar effective one body waveform model for spin-aligned black
  hole binaries}.
\newblock {\em Phys. Rev. D}, 102(2):024077, 2020.

\bibitem{Riemenschneider:2021ppj}
Gunnar Riemenschneider, Piero Rettegno, Matteo Breschi, Angelica Albertini,
  Rossella Gamba, Sebastiano Bernuzzi, and Alessandro Nagar.
\newblock {Assessment of consistent next-to-quasicircular corrections and
  postadiabatic approximation in effective-one-body multipolar waveforms for
  binary black hole coalescences}.
\newblock {\em Phys. Rev. D}, 104(10):104045, 2021.

\bibitem{Scheel:2014ina}
Mark~A. Scheel, Matthew Giesler, Daniel~A. Hemberger, Geoffrey Lovelace, Kevin
  Kuper, Michael Boyle, B.~Szil\'agyi, and Lawrence~E. Kidder.
\newblock {Improved methods for simulating nearly extremal binary black holes}.
\newblock {\em Class. Quant. Grav.}, 32(10):105009, 2015.

\end{thebibliography}


\clearpage

\begin{thebibliography}{10}
\expandafter\ifx\csname url\endcsname\relax
  \def\url#1{\texttt{#1}}\fi
\expandafter\ifx\csname urlprefix\endcsname\relax\def\urlprefix{URL }\fi
\providecommand{\bibinfo}[2]{#2}
\providecommand{\eprint}[2][]{\url{#2}}

\bibitem{TheLIGOScientific:2014jea}
\bibinfo{author}{Aasi, J.} \emph{et~al.}
\newblock \bibinfo{title}{{Advanced LIGO}}.
\newblock \emph{\bibinfo{journal}{Class. Quant. Grav.}}
  \textbf{\bibinfo{volume}{32}}, \bibinfo{pages}{074001}
  (\bibinfo{year}{2015}).
\newblock \eprint{1411.4547}.

\bibitem{TheVirgo:2014hva}
\bibinfo{author}{Acernese, F.} \emph{et~al.}
\newblock \bibinfo{title}{{Advanced Virgo: a second-generation interferometric
  gravitational wave detector}}.
\newblock \emph{\bibinfo{journal}{Class. Quant. Grav.}}
  \textbf{\bibinfo{volume}{32}}, \bibinfo{pages}{024001}
  (\bibinfo{year}{2015}).
\newblock \eprint{1408.3978}.

\bibitem{LIGOScientific:2021djp}
\bibinfo{author}{Abbott, R.} \emph{et~al.}
\newblock \bibinfo{title}{{GWTC-3: Compact Binary Coalescences Observed by LIGO
  and Virgo During the Second Part of the Third Observing Run}}
  (\bibinfo{year}{2021}).
\newblock \eprint{2111.03606}.

\bibitem{Abbott:2020tfl}
\bibinfo{author}{Abbott, R.} \emph{et~al.}
\newblock \bibinfo{title}{{GW190521: A Binary Black Hole Merger with a Total
  Mass of 150\,\,M\ensuremath{\odot}}}.
\newblock \emph{\bibinfo{journal}{Phys. Rev. Lett.}}
  \textbf{\bibinfo{volume}{125}}, \bibinfo{pages}{101102}
  (\bibinfo{year}{2020}).
\newblock \eprint{2009.01075}.

\bibitem{Abbott:2020mjq}
\bibinfo{author}{Abbott, R.} \emph{et~al.}
\newblock \bibinfo{title}{{Properties and astrophysical implications of the 150
  Msun binary black hole merger GW190521}}.
\newblock \emph{\bibinfo{journal}{Astrophys. J. Lett.}}
  \textbf{\bibinfo{volume}{900}}, \bibinfo{pages}{L13} (\bibinfo{year}{2020}).
\newblock \eprint{2009.01190}.

\bibitem{LIGOScientific:2020kqk}
\bibinfo{author}{Abbott, R.} \emph{et~al.}
\newblock \bibinfo{title}{{Population Properties of Compact Objects from the
  Second LIGO-Virgo Gravitational-Wave Transient Catalog}}.
\newblock \emph{\bibinfo{journal}{Astrophys. J. Lett.}}
  \textbf{\bibinfo{volume}{913}}, \bibinfo{pages}{L7} (\bibinfo{year}{2021}).
\newblock \eprint{2010.14533}.

\bibitem{Gonzalez:2020xah}
\bibinfo{author}{Gonz\'alez, E.} \emph{et~al.}
\newblock \bibinfo{title}{{Intermediate-mass Black Holes from High Massive-star
  Binary Fractions in Young Star Clusters}}.
\newblock \emph{\bibinfo{journal}{Astrophys. J. Lett.}}
  \textbf{\bibinfo{volume}{908}}, \bibinfo{pages}{L29} (\bibinfo{year}{2021}).
\newblock \eprint{2012.10497}.

\bibitem{Belczynski:2020bca}
\bibinfo{author}{Belczynski, K.}
\newblock \bibinfo{title}{{The most ordinary formation of the most unusual
  double black hole merger}}.
\newblock \emph{\bibinfo{journal}{Astrophys. J. Lett.}}
  \textbf{\bibinfo{volume}{905}}, \bibinfo{pages}{L15} (\bibinfo{year}{2020}).
\newblock \eprint{2009.13526}.

\bibitem{Mapelli:2021syv}
\bibinfo{author}{Mapelli, M.} \emph{et~al.}
\newblock \bibinfo{title}{{Hierarchical black hole mergers in young, globular
  and nuclear star clusters: the effect of metallicity, spin and cluster
  properties}}.
\newblock \emph{\bibinfo{journal}{Mon.\ Not.\ R.\ Astron.\ Soc.}}
  \textbf{\bibinfo{volume}{505}}, \bibinfo{pages}{339--358}
  (\bibinfo{year}{2021}).
\newblock \eprint{2103.05016}.

\bibitem{Sedda:2021abh}
\bibinfo{author}{Sedda, M.~A.} \emph{et~al.}
\newblock \bibinfo{title}{{Breaching the limit: formation of GW190521-like and
  IMBH mergers in young massive clusters}}  (\bibinfo{year}{2021}).
\newblock \eprint{2105.07003}.

\bibitem{Tagawa:2021ofj}
\bibinfo{author}{Tagawa, H.}, \bibinfo{author}{Haiman, Z.},
  \bibinfo{author}{Bartos, I.}, \bibinfo{author}{Kocsis, B.} \&
  \bibinfo{author}{Omukai, K.}
\newblock \bibinfo{title}{{Signatures of Hierarchical Mergers in Black Hole
  Spin and Mass distribution}}  (\bibinfo{year}{2021}).
\newblock \eprint{2104.09510}.

\bibitem{DallAmico:2021umv}
\bibinfo{author}{Dall'Amico, M.} \emph{et~al.}
\newblock \bibinfo{title}{{GW190521 formation via three-body encounters in
  young massive star clusters}}  (\bibinfo{year}{2021}).
\newblock \eprint{2105.12757}.

\bibitem{2020ApJ...902L..26F}
\bibinfo{author}{{Fragione}, G.}, \bibinfo{author}{{Loeb}, A.} \&
  \bibinfo{author}{{Rasio}, F.~A.}
\newblock \bibinfo{title}{{On the Origin of GW190521-like Events from Repeated
  Black Hole Mergers in Star Clusters}}.
\newblock \emph{\bibinfo{journal}{Astrophys. J. Lett.}}
  \textbf{\bibinfo{volume}{902}}, \bibinfo{pages}{L26} (\bibinfo{year}{2020}).
\newblock \eprint{2009.05065}.

\bibitem{2021arXiv210704639F}
\bibinfo{author}{{Fragione}, G.}, \bibinfo{author}{{Kocsis}, B.},
  \bibinfo{author}{{Rasio}, F.~A.} \& \bibinfo{author}{{Silk}, J.}
\newblock \bibinfo{title}{{Repeated mergers, mass-gap black holes, and
  formation of intermediate-mass black holes in nuclear star clusters}}.
\newblock \emph{\bibinfo{journal}{arXiv e-prints}}
  \bibinfo{pages}{arXiv:2107.04639} (\bibinfo{year}{2021}).
\newblock \eprint{2107.04639}.

\bibitem{Gayathri:2020coq}
\bibinfo{author}{Gayathri, V.} \emph{et~al.}
\newblock \bibinfo{title}{{Eccentricity estimate for black hole mergers with
  numerical relativity simulations}}.
\newblock \emph{\bibinfo{journal}{Nature Astron.}}
  \textbf{\bibinfo{volume}{6}}, \bibinfo{pages}{344--349}
  (\bibinfo{year}{2022}).
\newblock \eprint{2009.05461}.

\bibitem{Romero-Shaw:2020thy}
\bibinfo{author}{Romero-Shaw, I.~M.}, \bibinfo{author}{Lasky, P.~D.},
  \bibinfo{author}{Thrane, E.} \& \bibinfo{author}{Bustillo, J.~C.}
\newblock \bibinfo{title}{{GW190521: orbital eccentricity and signatures of
  dynamical formation in a binary black hole merger signal}}.
\newblock \emph{\bibinfo{journal}{Astrophys. J. Lett.}}
  \textbf{\bibinfo{volume}{903}}, \bibinfo{pages}{L5} (\bibinfo{year}{2020}).
\newblock \eprint{2009.04771}.

\bibitem{CalderonBustillo:2020odh}
\bibinfo{author}{Bustillo, J.~C.}, \bibinfo{author}{Sanchis-Gual, N.},
  \bibinfo{author}{Torres-Forn\'e, A.} \& \bibinfo{author}{Font, J.~A.}
\newblock \bibinfo{title}{{Confusing Head-On Collisions with Precessing
  Intermediate-Mass Binary Black Hole Mergers}}.
\newblock \emph{\bibinfo{journal}{Phys. Rev. Lett.}}
  \textbf{\bibinfo{volume}{126}}, \bibinfo{pages}{201101}
  (\bibinfo{year}{2021}).
\newblock \eprint{2009.01066}.

\bibitem{CalderonBustillo:2020srq}
\bibinfo{author}{Bustillo, J.~C.} \emph{et~al.}
\newblock \bibinfo{title}{{GW190521 as a Merger of Proca Stars: A Potential New
  Vector Boson of $8.7\times 10^{-13}$ eV}}.
\newblock \emph{\bibinfo{journal}{Phys. Rev. Lett.}}
  \textbf{\bibinfo{volume}{126}}, \bibinfo{pages}{081101}
  (\bibinfo{year}{2021}).
\newblock \eprint{2009.05376}.

\bibitem{Shibata:2021sau}
\bibinfo{author}{Shibata, M.}, \bibinfo{author}{Kiuchi, K.},
  \bibinfo{author}{Fujibayashi, S.} \& \bibinfo{author}{Sekiguchi, Y.}
\newblock \bibinfo{title}{{Alternative possibility of GW190521: Gravitational
  waves from high-mass black hole-disk systems}}.
\newblock \emph{\bibinfo{journal}{Phys. Rev. D}}
  \textbf{\bibinfo{volume}{103}}, \bibinfo{pages}{063037}
  (\bibinfo{year}{2021}).
\newblock \eprint{2101.05440}.

\bibitem{Nitz:2020mga}
\bibinfo{author}{Nitz, A.~H.} \& \bibinfo{author}{Capano, C.~D.}
\newblock \bibinfo{title}{{GW190521 may be an intermediate mass ratio
  inspiral}}.
\newblock \emph{\bibinfo{journal}{Astrophys. J. Lett.}}
  \textbf{\bibinfo{volume}{907}}, \bibinfo{pages}{L9} (\bibinfo{year}{2021}).
\newblock \eprint{2010.12558}.

\bibitem{Estelles:2021jnz}
\bibinfo{author}{Estell\'es, H.} \emph{et~al.}
\newblock \bibinfo{title}{{A detailed analysis of GW190521 with
  phenomenological waveform models}}  (\bibinfo{year}{2021}).
\newblock \eprint{2105.06360}.

\bibitem{East:2012xq}
\bibinfo{author}{East, W.~E.}, \bibinfo{author}{McWilliams, S.~T.},
  \bibinfo{author}{Levin, J.} \& \bibinfo{author}{Pretorius, F.}
\newblock \bibinfo{title}{{Observing complete gravitational wave signals from
  dynamical capture binaries}}.
\newblock \emph{\bibinfo{journal}{Phys. Rev.}} \textbf{\bibinfo{volume}{D87}},
  \bibinfo{pages}{043004} (\bibinfo{year}{2013}).
\newblock \eprint{1212.0837}.

\bibitem{Gold:2012tk}
\bibinfo{author}{Gold, R.} \& \bibinfo{author}{Br{\"u}gmann, B.}
\newblock \bibinfo{title}{{Eccentric black hole mergers and zoom-whirl behavior
  from elliptic inspirals to hyperbolic encounters}}.
\newblock \emph{\bibinfo{journal}{Phys. Rev.}} \textbf{\bibinfo{volume}{D88}},
  \bibinfo{pages}{064051} (\bibinfo{year}{2013}).
\newblock \eprint{1209.4085}.

\bibitem{Loutrel:2020kmm}
\bibinfo{author}{Loutrel, N.}
\newblock \bibinfo{title}{{Repeated Bursts: Gravitational Waves from Highly
  Eccentric Binaries}}  (\bibinfo{year}{2020}).
\newblock \eprint{2009.11332}.

\bibitem{Rasskazov:2019gjw}
\bibinfo{author}{Rasskazov, A.} \& \bibinfo{author}{Kocsis, B.}
\newblock \bibinfo{title}{{The rate of stellar mass black hole scattering in
  galactic nuclei}}.
\newblock \emph{\bibinfo{journal}{Astrophys. J.}}
  \textbf{\bibinfo{volume}{881}}, \bibinfo{pages}{20} (\bibinfo{year}{2019}).
\newblock \eprint{1902.03242}.

\bibitem{Tagawa:2019osr}
\bibinfo{author}{Tagawa, H.}, \bibinfo{author}{Haiman, Z.} \&
  \bibinfo{author}{Kocsis, B.}
\newblock \bibinfo{title}{{Formation and Evolution of Compact Object Binaries
  in AGN Disks}}.
\newblock \emph{\bibinfo{journal}{Astrophys. J.}}
  \textbf{\bibinfo{volume}{898}}, \bibinfo{pages}{25} (\bibinfo{year}{2020}).
\newblock \eprint{1912.08218}.

\bibitem{Rodriguez:2018pss}
\bibinfo{author}{Rodriguez, C.~L.} \emph{et~al.}
\newblock \bibinfo{title}{{Post-Newtonian Dynamics in Dense Star Clusters:
  Formation, Masses, and Merger Rates of Highly-Eccentric Black Hole
  Binaries}}.
\newblock \emph{\bibinfo{journal}{Phys. Rev. D}} \textbf{\bibinfo{volume}{98}},
  \bibinfo{pages}{123005} (\bibinfo{year}{2018}).
\newblock \eprint{1811.04926}.

\bibitem{Mukherjee:2020hnm}
\bibinfo{author}{Mukherjee, S.}, \bibinfo{author}{Mitra, S.} \&
  \bibinfo{author}{Chatterjee, S.}
\newblock \bibinfo{title}{{Detectability of hyperbolic encounters of compact
  stars with ground-based gravitational waves detectors}}
  (\bibinfo{year}{2020}).
\newblock \eprint{2010.00916}.

\bibitem{Mandel:2021smh}
\bibinfo{author}{Mandel, I.} \& \bibinfo{author}{Broekgaarden, F.~S.}
\newblock \bibinfo{title}{{Rates of Compact Object Coalescences}}
  (\bibinfo{year}{2021}).
\newblock \eprint{2107.14239}.

\bibitem{Chiaramello:2020ehz}
\bibinfo{author}{Chiaramello, D.} \& \bibinfo{author}{Nagar, A.}
\newblock \bibinfo{title}{{Faithful analytical effective-one-body waveform
  model for spin-aligned, moderately eccentric, coalescing black hole
  binaries}}.
\newblock \emph{\bibinfo{journal}{Phys. Rev. D}}
  \textbf{\bibinfo{volume}{101}}, \bibinfo{pages}{101501}
  (\bibinfo{year}{2020}).
\newblock \eprint{2001.11736}.

\bibitem{Nagar:2020xsk}
\bibinfo{author}{Nagar, A.}, \bibinfo{author}{Rettegno, P.},
  \bibinfo{author}{Gamba, R.} \& \bibinfo{author}{Bernuzzi, S.}
\newblock \bibinfo{title}{{Effective-one-body waveforms from dynamical captures
  in black hole binaries}}.
\newblock \emph{\bibinfo{journal}{Phys. Rev. D}}
  \textbf{\bibinfo{volume}{103}}, \bibinfo{pages}{064013}
  (\bibinfo{year}{2021}).
\newblock \eprint{2009.12857}.

\bibitem{Buonanno:1998gg}
\bibinfo{author}{Buonanno, A.} \& \bibinfo{author}{Damour, T.}
\newblock \bibinfo{title}{{Effective one-body approach to general relativistic
  two-body dynamics}}.
\newblock \emph{\bibinfo{journal}{Phys. Rev.}} \textbf{\bibinfo{volume}{D59}},
  \bibinfo{pages}{084006} (\bibinfo{year}{1999}).
\newblock \eprint{gr-qc/9811091}.

\bibitem{Blanchet:2013haa}
\bibinfo{author}{Blanchet, L.}
\newblock \bibinfo{title}{{Gravitational Radiation from Post-Newtonian Sources
  and Inspiralling Compact Binaries}}.
\newblock \emph{\bibinfo{journal}{Living Rev. Relativity}}
  \textbf{\bibinfo{volume}{17}}, \bibinfo{pages}{2} (\bibinfo{year}{2014}).
\newblock \eprint{1310.1528}.

\bibitem{Schafer:2018kuf}
\bibinfo{author}{Schaefer, G.} \& \bibinfo{author}{Jaranowski, P.}
\newblock \bibinfo{title}{{Hamiltonian formulation of general relativity and
  post-Newtonian dynamics of compact binaries}}.
\newblock \emph{\bibinfo{journal}{Living Rev. Rel.}}
  \textbf{\bibinfo{volume}{21}}, \bibinfo{pages}{7} (\bibinfo{year}{2018}).
\newblock \eprint{1805.07240}.

\bibitem{Nagar:2020pcj}
\bibinfo{author}{Nagar, A.}, \bibinfo{author}{Riemenschneider, G.},
  \bibinfo{author}{Pratten, G.}, \bibinfo{author}{Rettegno, P.} \&
  \bibinfo{author}{Messina, F.}
\newblock \bibinfo{title}{{Multipolar effective one body waveform model for
  spin-aligned black hole binaries}}.
\newblock \emph{\bibinfo{journal}{Phys. Rev. D}}
  \textbf{\bibinfo{volume}{102}}, \bibinfo{pages}{024077}
  (\bibinfo{year}{2020}).
\newblock \eprint{2001.09082}.

\bibitem{Pretorius:2007jn}
\bibinfo{author}{Pretorius, F.} \& \bibinfo{author}{Khurana, D.}
\newblock \bibinfo{title}{{Black hole mergers and unstable circular orbits}}.
\newblock \emph{\bibinfo{journal}{Class.Quant.Grav.}}
  \textbf{\bibinfo{volume}{24}}, \bibinfo{pages}{S83--S108}
  (\bibinfo{year}{2007}).
\newblock \eprint{gr-qc/0702084}.

\bibitem{Healy:2009zm}
\bibinfo{author}{Healy, J.}, \bibinfo{author}{Levin, J.} \&
  \bibinfo{author}{Shoemaker, D.}
\newblock \bibinfo{title}{{Zoom-Whirl Orbits in Black Hole Binaries}}.
\newblock \emph{\bibinfo{journal}{Phys.Rev.Lett.}}
  \textbf{\bibinfo{volume}{103}}, \bibinfo{pages}{131101}
  (\bibinfo{year}{2009}).
\newblock \eprint{0907.0671}.

\bibitem{Sperhake:2009jz}
\bibinfo{author}{Sperhake, U.} \emph{et~al.}
\newblock \bibinfo{title}{{Cross section, final spin and zoom-whirl behavior in
  high-energy black hole collisions}}.
\newblock \emph{\bibinfo{journal}{Phys.Rev.Lett.}}
  \textbf{\bibinfo{volume}{103}}, \bibinfo{pages}{131102}
  (\bibinfo{year}{2009}).
\newblock \eprint{0907.1252}.

\bibitem{Damour:2014afa}
\bibinfo{author}{Damour, T.} \emph{et~al.}
\newblock \bibinfo{title}{{Strong-Field Scattering of Two Black Holes: Numerics
  Versus Analytics}}  (\bibinfo{year}{2014}).
\newblock \eprint{1402.7307}.

\bibitem{Hopper:2022rwo}
\bibinfo{author}{Hopper, S.}, \bibinfo{author}{Nagar, A.} \&
  \bibinfo{author}{Rettegno, P.}
\newblock \bibinfo{title}{{Strong-field scattering of two spinning black holes:
  Numerics versus Analytics}}  (\bibinfo{year}{2022}).
\newblock \eprint{2204.10299}.

\bibitem{Aghanim:2018eyx}
\bibinfo{author}{Aghanim, N.} \emph{et~al.}
\newblock \bibinfo{title}{{Planck 2018 results. VI. Cosmological parameters}}.
\newblock \emph{\bibinfo{journal}{Astron. Astrophys.}}
  \textbf{\bibinfo{volume}{641}}, \bibinfo{pages}{A6} (\bibinfo{year}{2020}).
\newblock \eprint{1807.06209}.

\bibitem{Varma:2019csw}
\bibinfo{author}{Varma, V.} \emph{et~al.}
\newblock \bibinfo{title}{{Surrogate models for precessing binary black hole
  simulations with unequal masses}}.
\newblock \emph{\bibinfo{journal}{Phys. Rev. Research.}}
  \textbf{\bibinfo{volume}{1}}, \bibinfo{pages}{033015} (\bibinfo{year}{2019}).
\newblock \eprint{1905.09300}.

\bibitem{Akcay:2020qrj}
\bibinfo{author}{Akcay, S.}, \bibinfo{author}{Gamba, R.} \&
  \bibinfo{author}{Bernuzzi, S.}
\newblock \bibinfo{title}{{A hybrid post-Newtonian -- effective-one-body scheme
  for spin-precessing compact-binary waveforms}}.
\newblock \emph{\bibinfo{journal}{Phys. Rev. D}}
  \textbf{\bibinfo{volume}{103}}, \bibinfo{pages}{024014}
  (\bibinfo{year}{2021}).
\newblock \eprint{2005.05338}.

\bibitem{Gamba:2021ydi}
\bibinfo{author}{Gamba, R.}, \bibinfo{author}{Ak\c{c}ay, S.},
  \bibinfo{author}{Bernuzzi, S.} \& \bibinfo{author}{Williams, J.}
\newblock \bibinfo{title}{{Effective-one-body waveforms for precessing
  coalescing compact binaries with post-Newtonian twist}}.
\newblock \emph{\bibinfo{journal}{Phys. Rev. D}}
  \textbf{\bibinfo{volume}{106}}, \bibinfo{pages}{024020}
  (\bibinfo{year}{2022}).
\newblock \eprint{2111.03675}.

\bibitem{Breschi:2021wzr}
\bibinfo{author}{Breschi, M.}, \bibinfo{author}{Gamba, R.} \&
  \bibinfo{author}{Bernuzzi, S.}
\newblock \bibinfo{title}{{Bayesian inference of multimessenger astrophysical
  data: Methods and applications to gravitational waves}}.
\newblock \emph{\bibinfo{journal}{Phys. Rev. D}}
  \textbf{\bibinfo{volume}{104}}, \bibinfo{pages}{042001}
  (\bibinfo{year}{2021}).
\newblock \eprint{2102.00017}.

\bibitem{Nagar:2021gss}
\bibinfo{author}{Nagar, A.}, \bibinfo{author}{Bonino, A.} \&
  \bibinfo{author}{Rettegno, P.}
\newblock \bibinfo{title}{{Effective one-body multipolar waveform model for
  spin-aligned, quasicircular, eccentric, hyperbolic black hole binaries}}.
\newblock \emph{\bibinfo{journal}{Phys. Rev. D}}
  \textbf{\bibinfo{volume}{103}}, \bibinfo{pages}{104021}
  (\bibinfo{year}{2021}).
\newblock \eprint{2101.08624}.

\bibitem{Healy:2022wdn}
\bibinfo{author}{Healy, J.} \& \bibinfo{author}{Lousto, C.~O.}
\newblock \bibinfo{title}{{The Fourth RIT binary black hole simulations
  catalog: Extension to Eccentric Orbits}}  (\bibinfo{year}{2022}).
\newblock \eprint{2202.00018}.

\bibitem{Harms:2014dqa}
\bibinfo{author}{Harms, E.}, \bibinfo{author}{Bernuzzi, S.},
  \bibinfo{author}{Nagar, A.} \& \bibinfo{author}{Zenginoglu, A.}
\newblock \bibinfo{title}{{A new gravitational wave generation algorithm for
  particle perturbations of the Kerr spacetime}}.
\newblock \emph{\bibinfo{journal}{Class.Quant.Grav.}}
  \textbf{\bibinfo{volume}{31}}, \bibinfo{pages}{245004}
  (\bibinfo{year}{2014}).
\newblock \eprint{1406.5983}.

\bibitem{Albanesi:2021rby}
\bibinfo{author}{Albanesi, S.}, \bibinfo{author}{Nagar, A.} \&
  \bibinfo{author}{Bernuzzi, S.}
\newblock \bibinfo{title}{{Effective one-body model for extreme-mass-ratio
  spinning binaries on eccentric equatorial orbits: Testing radiation reaction
  and waveform}}.
\newblock \emph{\bibinfo{journal}{Phys. Rev. D}}
  \textbf{\bibinfo{volume}{104}}, \bibinfo{pages}{024067}
  (\bibinfo{year}{2021}).
\newblock \eprint{2104.10559}.

\bibitem{Abbott:2019ebz}
\bibinfo{author}{Abbott, R.} \emph{et~al.}
\newblock \bibinfo{title}{{Open data from the first and second observing runs
  of Advanced LIGO and Advanced Virgo}}  (\bibinfo{year}{2019}).
\newblock \eprint{1912.11716}.

\bibitem{Speagle:2020}
\bibinfo{author}{Speagle, J.~S.}
\newblock \bibinfo{title}{dynesty: a dynamic nested sampling package for
  estimating bayesian posteriors and evidences}.
\newblock \emph{\bibinfo{journal}{Monthly Notices of the Royal Astronomical
  Society}} \textbf{\bibinfo{volume}{493}}, \bibinfo{pages}{3132?3158}
  (\bibinfo{year}{2020}).
\newblock \urlprefix\url{http://dx.doi.org/10.1093/mnras/staa278}.

\bibitem{Cao:2017ndf}
\bibinfo{author}{Cao, Z.} \& \bibinfo{author}{Han, W.-B.}
\newblock \bibinfo{title}{{Waveform model for an eccentric binary black hole
  based on the effective-one-body-numerical-relativity formalism}}.
\newblock \emph{\bibinfo{journal}{Phys. Rev.}} \textbf{\bibinfo{volume}{D96}},
  \bibinfo{pages}{044028} (\bibinfo{year}{2017}).
\newblock \eprint{1708.00166}.

\end{thebibliography}
\end{document}